\documentclass[a4paper,fleqn,usenatbib,useAMS]{mnras}
\usepackage{graphicx}	
\usepackage{amsmath}	
\usepackage{amssymb}	
\usepackage{multicol}        
\usepackage{bm}		
\usepackage{pdflscape}	
\usepackage[T1]{fontenc}
\usepackage{ae,aecompl}
\usepackage{txfonts}
\usepackage{threeparttable}
\usepackage{tikz}
\usetikzlibrary{shapes.geometric, arrows}

\tikzstyle{stardust's} = [rectangle, rounded corners, minimum width=3cm, minimum height=1cm,text centered, draw=black, fill=red!30]
\tikzstyle{io} = [trapezium, trapezium left angle=70, trapezium right angle=110, minimum width=3cm, minimum height=1cm, text centered, draw=black, fill=blue!30]
\tikzstyle{process} = [rectangle, minimum width=4cm, minimum height=1cm, text centered, text width=3cm, draw=black, fill=orange!30]
\tikzstyle{decision} = [diamond, minimum width=3cm, minimum height=1cm, text centered, draw=black, fill=green!30]
\tikzstyle{arrow} = [thick,->,>=stealth]

\title[HIF--Photosphere Interaction]
{The stellar photosphere-hydrogen ionization front interaction in Classical Pulsators: a theoretical explanation for observed period-colour relations}
\author[Das et al.]{Susmita Das$^{1}$\thanks{E-mail: susmitadas130@gmail.com (SD); shashi.kanbur@oswego.edu (SMK)}, Shashi M. Kanbur$^{2}$, Earl P.~Bellinger$^{3}$\thanks{SAC Postdoctoral Fellow}, Anupam Bhardwaj$^{4}$\thanks{KIAA Postdoctoral Fellow},
\newauthor
 Harinder P. Singh$^{1}$, Brett Meerdink$^{2}$, Nicholas Proietti$^{2}$, Anthony Chalmers$^{2}$, and \and Ryan Jordan$^{2}$\\
\vspace{1pt}\\
1. Department of Physics \& Astrophysics, University of Delhi, Delhi 110007, India \\
2. Department of Physics, State University of New York Oswego, Oswego, NY 13126, USA\\
3. Stellar Astrophysics Centre, Department of Physics and Astronomy, Aarhus University, Aarhus 8000, Denmark \\
4. Kavli Institute for Astronomy and Astrophysics, Peking University, Yi He Yuan Lu 5, Hai Dian District, Beijing 100871, China
}

\begin{document}

\date{Accepted 2020 January 17. Received 2020 January 17; in original form 2019 November 15}

\pagerange{\pageref{firstpage}--\pageref{lastpage}} \pubyear{2019}

\maketitle

\label{firstpage}

\begin{abstract}
Period-colour and amplitude-colour (PCAC) relations can be used to probe both the hydrodynamics of outer envelope structure and evolutionary status of Cepheids and RR~Lyraes. In this work, we incorporate the PCAC relations for RR~Lyraes, BL~Her, W~Vir and classical Cepheids in a single unifying theory that involves the interaction of the hydrogen ionization front (HIF) and stellar photosphere and the theory of stellar evolution. PC relations for RR~Lyraes and classical Cepheids using OGLE-IV data are found to be consistent with this theory: RR~Lyraes have shallow/sloped relations at minimum/maximum light whilst long-period ($P>10$ days) Cepheids exhibit sloped/flat PC relations at minimum/maximum light. The differences in the PC relations for Cepheids and RR~Lyraes can be explained based on the relative location of the HIF and stellar photosphere which changes depending on their position on the HR diagram. We also extend our analysis of PCAC relations for type~II Cepheids in the Galactic bulge, LMC and SMC using OGLE-IV data. We find that BL~Her stars have sloped PC relations at maximum and minimum light similar to short-period ($P<10$ days) classical Cepheids. W~Vir stars exhibit sloped/flat PC relation at minimum/maximum light similar to long-period classical Cepheids. We also compute state-of-the-art 1D radiation hydrodynamic models of RR~Lyraes, BL~Her and classical Cepheids using the radial stellar pulsation code in MESA to further test these ideas theoretically and find that the models are generally consistent with this picture. We are thus able to explain PC relations at maximum and minimum light across a broad spectrum of variable star types.
\end{abstract}

\begin{keywords}
{stars: variables: Cepheids, stars: variables: RR Lyrae, stars: Population II, stars: evolution, Galaxy: bulge, galaxies: Magellanic Clouds}
\end{keywords}

\section{Introduction}

RR~Lyraes, classical Cepheids and Type~II Cepheids (T2Cs) are well-known classical pulsators in different evolutionary stages. While RR~Lyraes are horizontal-branch stars with helium-burning cores \citep{preston1964}, classical Cepheids are located on the blue loops in the HR diagram \citep{anderson2016}. 
T2Cs are old, low-mass (${\sim 0.5-0.6\;\text{M}_{\odot}}$), metal-poor pulsating variables with typical periods of about $1$ to $50$ days \citep{wallerstein2002,sandage2006}. T2Cs are fainter than the classical Cepheids but brighter than the RR Lyrae stars. They have traditionally been divided into three subclasses: BL~Her stars with periods between $1-4$ days, W~Vir stars with periods ranging from $4-20$ days and RV~Tau stars with periods longer than $20$ days \citep{soszynski2018}. We note here that RV~Tau variables are post AGB stars with typical masses of $\sim 0.7M_{\odot}$ \citep{tuchman1993, fokin1994}, higher than those for BL~Her and W~Vir stars. In addition, there is a fourth subclass, the peculiar W~Vir (pW~Vir) stars which are brighter and bluer than the W~Vir stars \citep{soszynski2008}. All these classical pulsating stars follow well-defined Period-Luminosity relations (PLRs) which make them excellent distance indicators \citep{muraveva2015, bhardwaj2016a, bhardwaj2017b, ripepi2017, subramanian2017, beaton2018}.

The PCAC relations for classical Cepheids and RR Lyraes have been extensively studied to understand the hydrodynamics in their outer envelopes \citep{simon1993,kanbur2004b, bhardwaj2014,ngeow2017, das2018}. RR~Lyrae stars are on the Horizontal Branch (HB) and evolve towards higher luminosities and cooler temperatures in the HR diagram. However, depending on the morphology of the evolutionary tracks, RR~Lyraes can evolve bluewards from a relatively red zero-age HB before turning redwards towards the AGB. Their period-colour (PC) relations have a particular structure: shallow/steep sloped relations at minimum/maximum light respectively. Classical Cepheids lie in a region of higher luminosities and cooler temperatures on the HR diagram with respect to the RR~Lyrae stars: their PC relations at minimum (steep slope) and maximum (shallow slope) light are the opposite of RR~Lyraes. \citet{kanbur2004b,bhardwaj2014,ngeow2017} and references therein contend that the changes in the PCAC relations are caused by an interaction of the stellar photosphere and hydrogen ionization front (HIF) in the outer envelope. The photosphere is taken as being at optical depth ${\tau}=2/3$ and the HIF is defined where the majority of hydrogen becomes ionized. However, PCAC relations for T2Cs at the maximum and minimum light have not been studied previously.

The classical picture of the theory of stellar evolution suggests that the progenitors of T2Cs evolve redward from the blue edge of the HB and may move up the Asymptotic Giant Branch (AGB) going through the BL~Her-type and W~Vir-type pulsation stages \citep{gingold1985, wallerstein2002}. However, recent work by \citet{gezer2015, groenewegen2017a, groenewegen2017b, manick2017} among others questions this classical picture and the origin and evolution of T2Cs is not as well-understood as was once thought to be. \citet{iwanek2018} confirmed BL~Her stars to be as old as RR~Lyraes while W~Vir stars could be a mixture of old and intermediate-age stars. While \citet{groenewegen2017a} found that BL~Her stars may be explained by low-mass stars ($\sim 0.5-0.6 M_{\odot}$) evolving off the zero-age HB, they state that the evolutionary status of W~Vir stars is unclear. This is because for the same luminosity, W~Vir stars have longer periods and thus, lower masses than their short period classical Cepheid counterparts. However, evolutionary tracks of $\sim 2.5-4 M_{\odot}$ stars cross the W~Vir region in the HR diagram. \citet{groenewegen2017a} suggested that the origin of W~Vir stars may have some relation to binarity. 

From the theoretical perspective, several linear and nonlinear convective models for RR~Lyraes, Cepheids and T2Cs have been computed to study their pulsation properties and PLRs \citep{fiorentino2007, marconi2007, smolec2012a, marconi2013, marconi2015, bono2016, smolec2016}. In the case of Cepheids and RR~Lyraes, their light curve structures have also been compared with observations \citep{bhardwaj2017a, das2018}. For T2Cs, \citet{marconi2007, smolec2016} have employed non-linear models to investigate their pulsation modes. While they pulsate primarily in the fundamental mode, double-mode and overtone-mode T2Cs have also been found by \citet{smolec2018} and \citet{soszynski2019}, respectively. Recently, \emph{Modules for Experiments in Stellar Astrophysics} \citep[MESA,][]{paxton2011,paxton2013,paxton2015,paxton2018,paxton2019} incorporated radial stellar pulsation modules which can be used to generate light curves of these classical pulsators. Our aim is to compare the observed PCAC relations for RR Lyraes, classical Cepheids and T2Cs with a self-consistent set of pulsation models and relate the empirical variations to changes in the envelope structure and evolutionary states of these variables.

There are several reasons for comparing observed and theoretical PCAC relations. Firstly, PC relations at mean light are important in a wider astrophysical context because through the period-luminosity-colour (PLC) relations, they influence the PLRs which is vital for the distance scale and non-CMB estimates of Hubble's constant \citep{beaton2016, riess2016, riess2019}. Traditionally, PC relations are studied at {\it mean} light \citep{tammann2003} but these relations are the average of the corresponding relations at various phase points during a pulsation cycle. Therefore in order to fully understand the physics behind these relations at mean light, it is important to study such relations during the pulsation cycle. Previous work has found small nonlinearities in the Cepheid PLRs \citep{bhardwaj2016b} which may occur due to the significant variations in the PC relation at maximum and minimum light.
Further, observed PCAC relations can be compared to theoretical PCAC relations obtained from extensive grids of theoretical pulsation models to constrain global stellar parameters \citep{bellinger2019} and hence constrain theories of stellar pulsation and evolution.

Here we focus on PC relations at maximum and minimum light and corresponding AC relations \citep[and references therein]{bhardwaj2014, das2018} because the behaviour at these phases has been shown to be strongly influenced by the interaction between the HIF and the stellar photosphere. In previous work, \citet{simon1993,kanbur1995,kanbur1996} investigated how such interactions can strongly influence the behavior of PC relations at maximum (classical Cepheids) and minimum light (RR Lyraes). Later \citet{kanbur2004b,kanbur2006, kanbur2007} modelled this interaction using 1D radiation hydrodynamic models of Cepheids and RR Lyraes. These authors found that the stellar photosphere and the HIF are not always co-moving during a pulsation cycle. When the two are ``engaged'' (that is, the photosphere lies at the base of the HIF or the distance between them is small) at low temperatures, the temperature of the photosphere takes on the properties of the temperature at which hydrogen ionizes. From Saha ionization equilibrium theory, this temperature is independent of density at low temperatures, leading to a PC relation that is flatter or more independent of global stellar pulsation properties. This is the situation, for example, for Cepheids at maximum light. At minimum light for Cepheids, the photosphere and the HIF are not engaged and the temperature of the photosphere is dependent on global stellar properties - hence we have a sloped PC relation. In the case of RR Lyraes, the photosphere and the HIF are engaged through the pulsation cycle \citep{kanbur1995, kanbur1996}. However the interaction is at low temperatures at minimum light and hence RR Lyraes exhibit a flat PC relation at minimum light. At maximum light, this interaction is at higher temperatures and the temperature at which hydrogen ionizes or the temperature of the stellar photosphere is strongly dependent on temperature and global stellar properties. Hence we have a definite slope for RRab PC relations at maximum light but a flat PC slope at minimum light. 

One measure of ``engagement'' between the photosphere and HIF is the distance between them in mass coordinates \citep[and references therein]{kanbur2007}. In this work, we study how this distance varies across classical Cepheids, RR Lyraes and T2Cs and across pulsation phases. We emphasize here that for the purpose of this paper, the locations of different types of variable stars on the HR diagram and thus, in different evolutionary states determine the relative location of HIF and photosphere. A higher $L/M$ ratio and/or a cooler effective temperature indicates greater distance between the HIF and the stellar photosphere \citep{kanbur1995, kanbur1996}. W~Vir stars lie in the same region of the HR diagram as classical Cepheids. The HIF-stellar photosphere interaction theory \citep{kanbur1995,kanbur1996} implies that as a consequence of stellar evolution and because of similar locations of W~Vir stars and classical Cepheids on the HR diagram, the PC relation obeyed by W~Vir stars should be similar to that for classical Cepheids. We also note that previous work has found some statistically significant differences in PCAC relations between the Galaxy, LMC and SMC, though the general nature (that is, flat PC relation at maximum/minimum for Cepheids/RR Lyraes) is preserved. Hence is is important to understand the effect of metallicity with a view to using such observed PC relations at multiple phases to constrain models. It is in this context that we study the PCAC properties of T2Cs in the Bulge and the Magellanic Clouds for the very first time, to the best of our knowledge using the best available data and pulsation codes and compare with results for Cepheids and RR Lyraes. Hence this project aims to find direct observational evidence supporting the HIF-stellar photosphere interaction in the outer envelopes of variable stars and to probe the PC relation across a broad spectrum of classical pulsating stars in different evolutionary stages.

The structure of this paper is as follows: The theoretical motivation of this project is discussed in Section~\ref{sec:theory} while Section~\ref{sec:data} describes the observational data used in this analysis and the Fourier decomposition technique. The extinction-corrected period-colour and amplitude-colour relations are briefly discussed in Section~\ref{sec:PCAC}. We study the distance between the HIF and the stellar photosphere of RR~Lyrae, BL~Her and classical Cepheid models computed using MESA in Section~\ref{sec:HIF-Ph}. Finally, we summarise the results of this study in Section~\ref{sec:results}.

\begin{table*}
\caption{The observed light curve data used in the present analysis with the number of stars available in each dataset. $N_{\text{Cep}}$ refers to the number of fundamental mode classical Cepheids while $N_{\text{T2C}}$ is the total number of type II Cepheids available in each band for each source.}
\centering
\begin{tabular}{c c c c c c c c c c}
\hline\hline
Source & Band & $N_{\text{Cep}}$ & Reference & $N_{\text{T2C}}$  & $N_{\text{BL~Her}}$ & $N_{\text{W~Vir}}$ & $N_{\text{pW~Vir}}$ & $N_{\text{RV~Tau}}$	& Reference\\
\hline \hline
& V & 5 & & 517 & 221 & 193 & 18 & 85 &\\[-1ex]
\raisebox{1ex}{Bulge (OGLE-IV)} & & & \raisebox{1ex}{\citet{soszynski2017}} & & & & & & \raisebox{1ex}{\citet{soszynski2017}}\\[-1ex]
& I & 30 & & 873 & 372 & 336 & 30  & 135 &\\[1ex]

& V & 2027 & & 243 & 76 & 92 & 24 & 51 &\\[-1ex]
\raisebox{1ex}{LMC (OGLE-IV)} & & & \raisebox{1ex}{\citet{soszynski2015}} & & & & & & \raisebox{1ex}{\citet{soszynski2018}}\\[-1ex]
& I & 2404 & & 253 & 80 & 96 & 25 & 52 &\\[1ex]

& V & 2422 & & 52 & 20 & 15 & 7 & 10 &\\[-1ex]
\raisebox{1ex}{SMC (OGLE-IV)} & & & \raisebox{1ex}{\citet{soszynski2015}} & & & & & & \raisebox{1ex}{\citet{soszynski2018}}\\[-1ex]
& I & 2687 & & 52 & 20 & 15 & 7 & 10 &\\[1ex]

\hline
\end{tabular}
\label{tab:Observed_Data}
\end{table*}

\section{Theoretical Framework}
\label{sec:theory}

\citet{kanbur1995} found that the HIF was further out in the mass distribution in RR Lyrae stars as compared to classical Cepheids, i.e., the HIF is closer to the surface of the star with a higher effective temperature. In fact, \citet{kanbur1995} noted that the HIF lies further in the mass distribution as $L/M$ increases and as $T_{\text{eff}}$ decreases. While this result may be expected because Cepheids are cooler than RR Lyraes, it is nevertheless instructive and important to be able to demonstrate it both computationally and theoretically. To physically justify this statement, let $\tau$ be the optical depth in terms of the Rosseland mean opacity. Then the temperature $T$ in the stellar envelope can be approximated by 
\begin{equation}
   T^4(\tau) 
   = 
   \frac{3}{4}\, 
   T_{\text{eff}}^4
   \left[ 
       \tau+q(\tau) 
   \right]
\end{equation}
where $T_{\text{eff}}$ is the effective temperature and $q(\tau)$ is the Hopf function, a monotonically increasing function of $\tau$ \citep[e.g.,][]{hopf1930}. We may consider $T_{\text{eff}}$ to be constant at $\sim 10000$K at the base of the HIF. Since $q(\tau)$ increases monotonically with $\tau$, the opacity of the HIF increases with decreasing effective temperature. Therefore, the HIF lies farther from the surface when the star is cooler. 

To investigate and contextualize this within the stellar evolution theory, we used MESA to calculate the evolution of a star with an initial mass of $0.83\;\text{M}_\odot$ and initial composition of ${Y=0.25,\; Z=0.001}$ and study the relative location of the HIF and the stellar photosphere as it moves off the horizontal branch. 
We used the \citet[][GS98]{grevesse1998} mixture for heavy-mass elements, nuclear reaction rates from the \emph{Nuclear Astrophysics Compilation of Reaction Rates} \citep[NACRE,][]{angulo1999}, and the \emph{Opacity Project at Livermore} (OPAL) equation of state and type-two opacities \citep{iglesias1996, rogers2002}. 
We treated convection using \citet{bohmvitense1958} mixing-length theory with a solar-calibrated mixing length (${\alpha_{\text{MLT}} \simeq 1.84}$). 
We calculated mass loss using Reimers' prescription with a mass loss rate of ${\eta=0.5}$ \citep{reimers1975,fadeyev2019}. 
We tracked the evolution from the pre-main sequence through the main sequence (MS), up the red giant branch (RGB), past the helium flash, onto the horizontal branch (HB), and finally onto the asymptotic giant branch (AGB). 

\begin{figure}
    \centering
    \includegraphics[width=\linewidth]{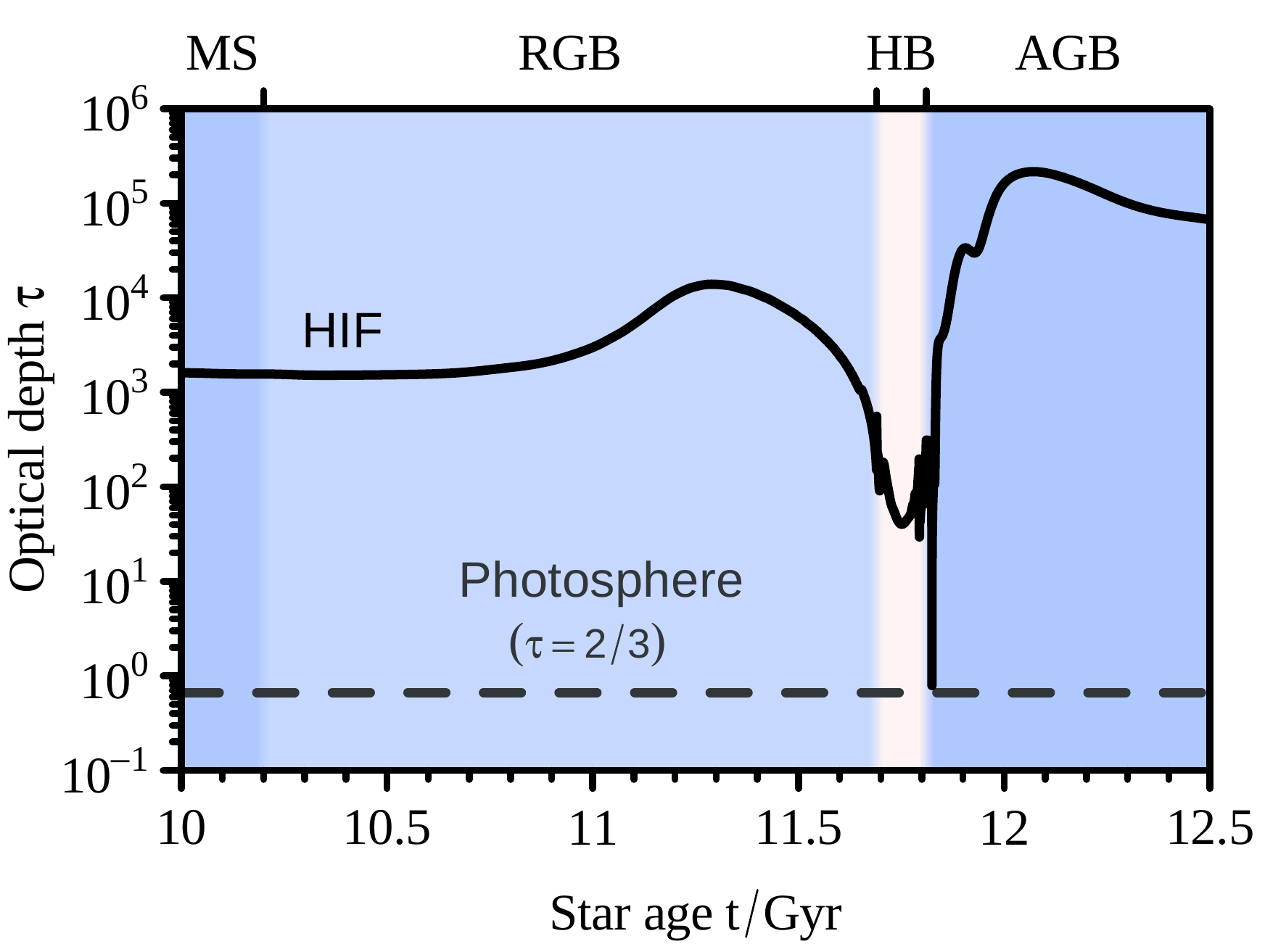}
    \caption{The location of the hydrogen ionization front compared to the photosphere as a function of stellar age for a low-mass star (${M\simeq 0.632\;\text{M}_\odot}$). 
        \label{fig:hif-tau-RRL}}
\end{figure}

\begin{figure*}
    \centering
    \includegraphics[width=\linewidth]{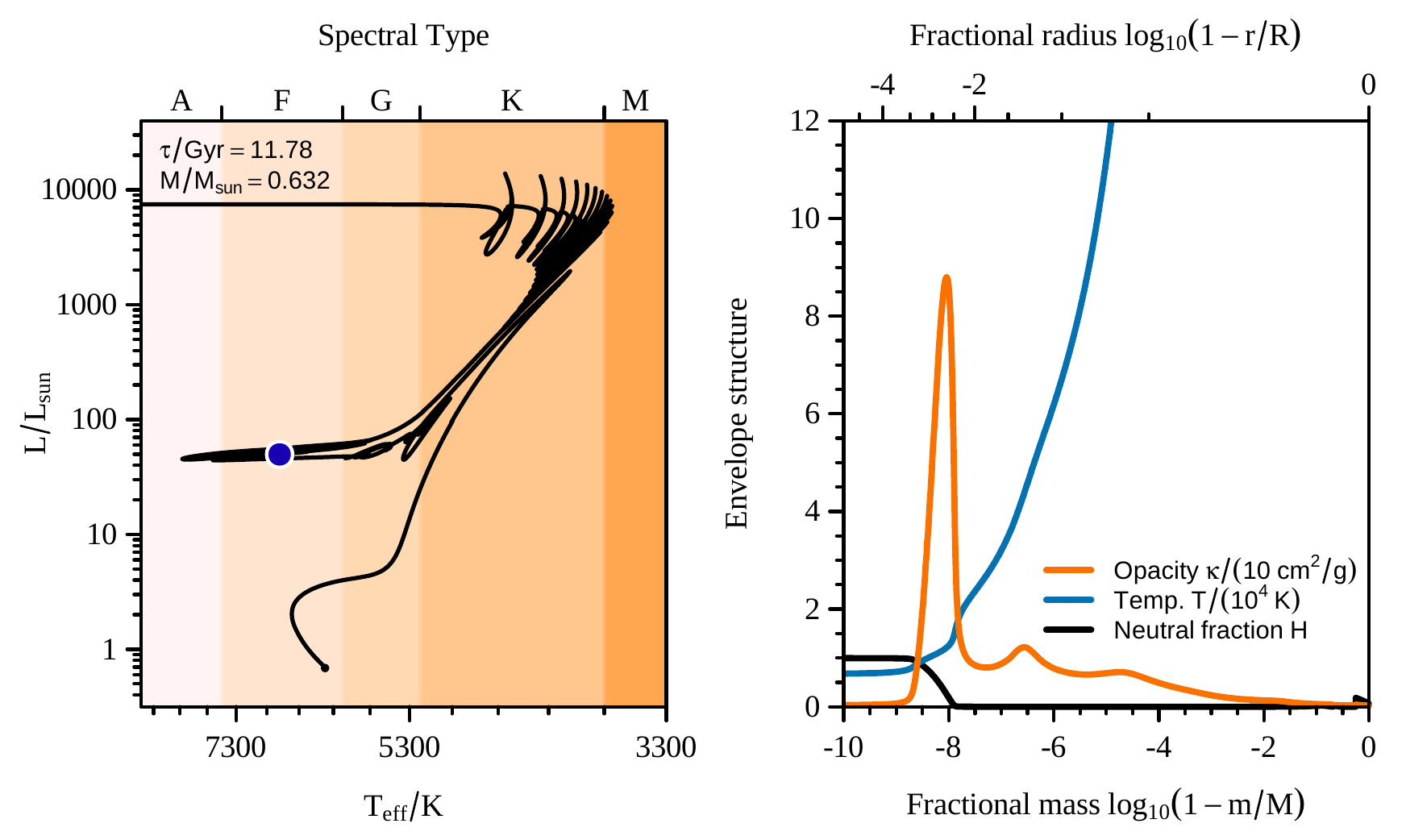}
    \caption{\emph{Left panel}: Hertzsprung-Russell diagram showing the theoretical evolution of a low-mass star with ${M_{\text{init}} = 0.83\;\text{M}_\odot}$. 
    The blue dot shows the position of an RR~Lyrae-type star on the horizontal branch. 
    The current mass and age of the model are indicated in the legend. 
    \emph{Right panel}: The thermodynamic envelope structure for the model indicated in the left panel. The hydrogen ionization front of an RR~Lyrae star at equilibrium is located near to the stellar photosphere. 
        \label{fig:hif-rrl}}
    \includegraphics[width=\linewidth]{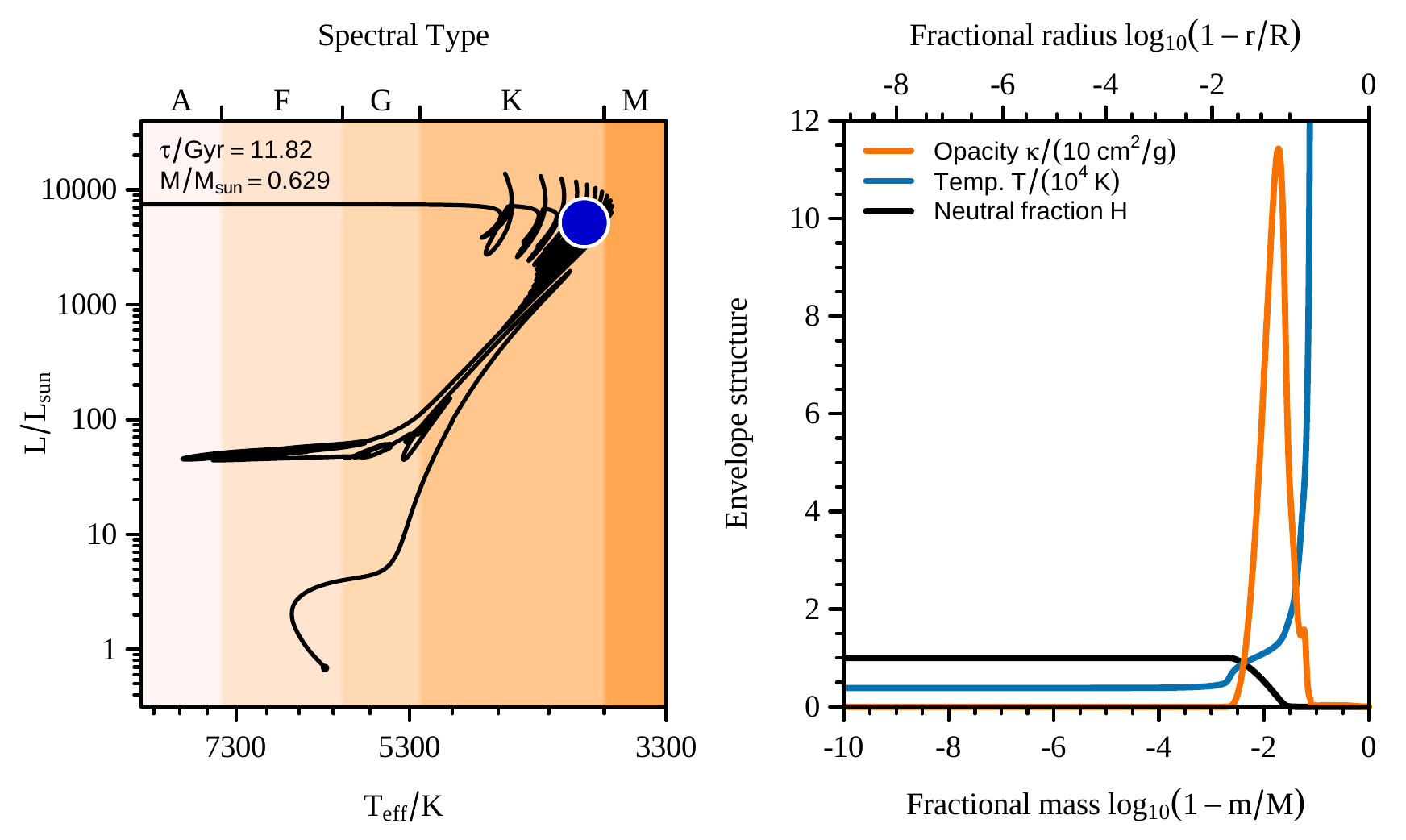}
    \caption{The same as Figure~\ref{fig:hif-rrl}, but at a later stage of evolution; now on the asymptotic giant branch. The hydrogen ionization front for an AGB star at equilibrium is located deeper into the stellar interior. 
        \label{fig:hif-agb}}
\end{figure*}

Fig.~\ref{fig:hif-tau-RRL} displays the location of the HIF compared to the photosphere as a function of stellar age. We note that this relative location or distance varies considerably during the evolutionary track.
During the HB phase of evolution, the HIF is located near to the stellar photosphere \citep{kanbur1995}. 
After the exhaustion of helium in the core, when the star moves off the HB, it moves to brighter
luminosities and cooler temperatures. That is, its $L/M$ ratio increases and its $T_{\text{eff}}$ decreases. Consequently, the HIF moves deeper into the mass distribution of the star. Figs.~\ref{fig:hif-tau-RRL}, \ref{fig:hif-rrl} and \ref{fig:hif-agb} display these arguments. Fig.~\ref{fig:hif-tau-RRL} presents the relative location of the photosphere and HIF at various
stages during the life of a $0.632M_{\odot}$ star. The star has an initial mass of $0.83M_{\odot}$ and only reaches this low mass at the advanced stages of evolution.
After the HB phase we see clearly that the HIF moves further away from the photosphere. Figs.~\ref{fig:hif-rrl} and \ref{fig:hif-agb} display this in greater detail.
The left panel of Fig.~\ref{fig:hif-rrl} presents the location of this star on the HB whilst the right panel displays its outer envelope structure - specifically
its temperature, opacity and ionized hydrogen structure. Fig.~\ref{fig:hif-agb} is the same but for a star on the AGB. We clearly see that as the star moves to the AGB, its HIF moves further into the mass distribution. 

\section{Data and methods}
\label{sec:data}
In order to confirm these proposed HIF-photosphere interactions, we now obtain the PCAC relations from observations for stars at various stages of stellar evolution. The optical ($VI$) light curves of the classical Cepheids and T2Cs in the Galactic bulge are taken from the OGLE-IV catalog \citep{soszynski2017} while those in Large Magellanic Cloud (LMC) and Small Magellanic Cloud (SMC) are taken from \citet{soszynski2015} for classical Cepheids and from \citet{soszynski2018} for T2Cs. The classification for the different subclasses of T2Cs is as provided by OGLE-IV. For our analysis, we choose the stars with well-sampled light curves having more than 30 data points in both $V$ and $I$ bands. The number of stars available for classical Cepheids and T2Cs is summarized in Table~\ref{tab:Observed_Data}. 

The photometric light curve data are fitted with the Fourier sine series \citep[see example,][]{deb2009,bhardwaj2015,das2018}:
\begin{equation}
m = m_0 + \sum_{k=1}^{N}A_k \sin(2 \pi kx+\phi_k),
\label{eq:fourier}
\end{equation}
where the phase $x$ is calculated as:
\begin{equation}
x=\frac{(t-t_0)}{P} - \textrm{int}\left(\frac{t-t_0}{P}\right),
\end{equation}
where $t_0$ is the epoch of maximum brightness and $P$ is the pulsation period. The order of the fit, $N$ is obtained using the Bart's criteria \citep{bart1982} by varying it from 4 to 8. 
From the Fourier-fitted light curves, we obtain colour at maximum, minimum and mean light as:

\begin{equation}
\begin{aligned}
(V-I)_{\max} &= V_{\max} - I_{\operatorname{phmax}},\\
(V-I)_{\min} &= V_{\min} - I_{\operatorname{phmin}},\\
(V-I)_{\operatorname{mean}} &= V_{\operatorname{mean}} - I_{\operatorname{mean}},
\end{aligned}
\end{equation}
where $I_{\operatorname{phmax}}$ and $I_{\operatorname{phmin}}$ correspond to the $I$-magnitude at the same phase as that of $V_{\max}$ and $V_{\min}$, respectively. The amplitude in $V$ is simply defined as the difference between the maximum and minimum of the magnitude in the $V$-band.

To correct for extinction in the classical Cepheids and T2Cs in the LMC and SMC, we obtain their colour excess ${E(V-I)}$ values using the reddening maps of \citet{haschke2011} and convert to their ${E(B-V)}$ values using $E(V-I) = 1.38 E(B-V) $\citep{tammann2003}. The extinction values are found by adopting the reddening law of \citet{cardelli1989}: $A_V = 3.32 E(B-V); A_I = 1.94 E(B-V)$\citep{schlegel1998}.

For the T2Cs in the Galactic bulge, we first use the extinction maps of \citet{gonzalez2012} and adopt the standard reddening law from \cite{cardelli1989} ($\frac{A_{K_s}}{A_V} = 0.114; \frac{A_I}{A_V} = 0.479$). Given the non-standard nature of the reddening law towards the central Galactic bulge \citep{popowski2000,udalski2003,nishiyama2009, nataf2013, matsunaga2013}, we also use the extinction calculator provided by \citet{nataf2013} to obtain the colour excess $E(V-I)$ and extinction $A_I$. This calculator primarily assumes the $E(J-K_s)$ measurements of \citet{gonzalez2012}. The $A_V$ values are then estimated by using $E(V-I)=A_V-A_I$. We only use the first method of extinction correction for the classical Cepheids in the Galactic bulge.

These extinction corrections are applied to the magnitudes and colours at minimum, mean, and maximum light during the pulsation cycle. We fit linear-regression models to extinction corrected PC and AC relations for different classes of pulsating stars. Throughout this paper, we employ an iterative 3$\sigma$ outlier clipping during the regression analysis and points beyond this threshold are considered outliers.

\begin{figure*}
\centering
\includegraphics[scale = 0.85]{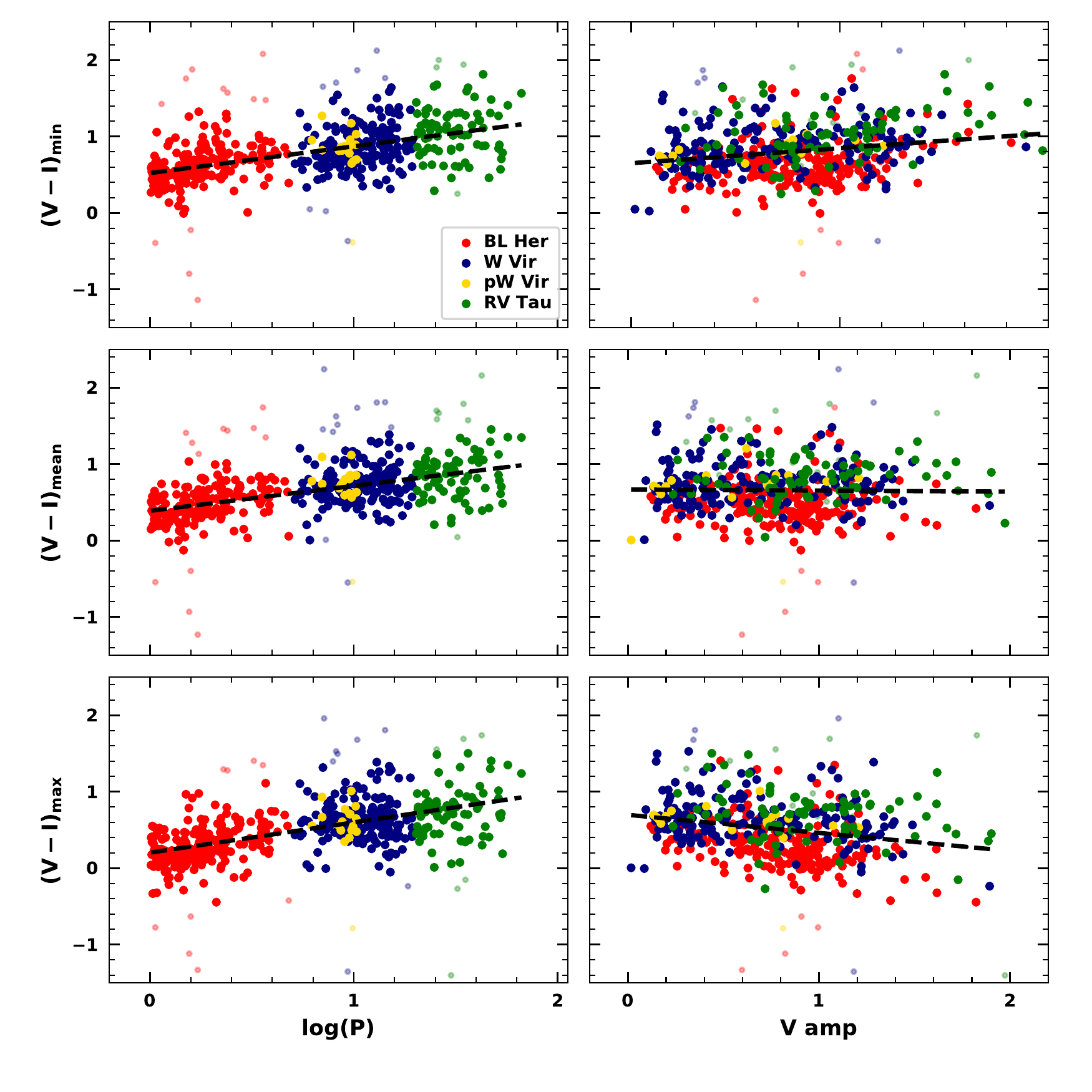}
\caption{PC and AC relations for T2Cs in the Bulge from OGLE-IV at minimum, mean and maximum light. The different sub-classes are shown in different colours. However, the dashed line is the best fit linear regression for T2Cs as a whole with the different sub-classes combined, after recursively removing the 3$\sigma$ outliers (depicted in smaller, shaded symbols). The PC and AC relations obtained for the sub-classes separately may be found in Table~\ref{tab:PCAC}.}
\label{fig:T2C_Bulge}
\end{figure*}

\begin{figure*}
\centering
\includegraphics[scale = 0.85]{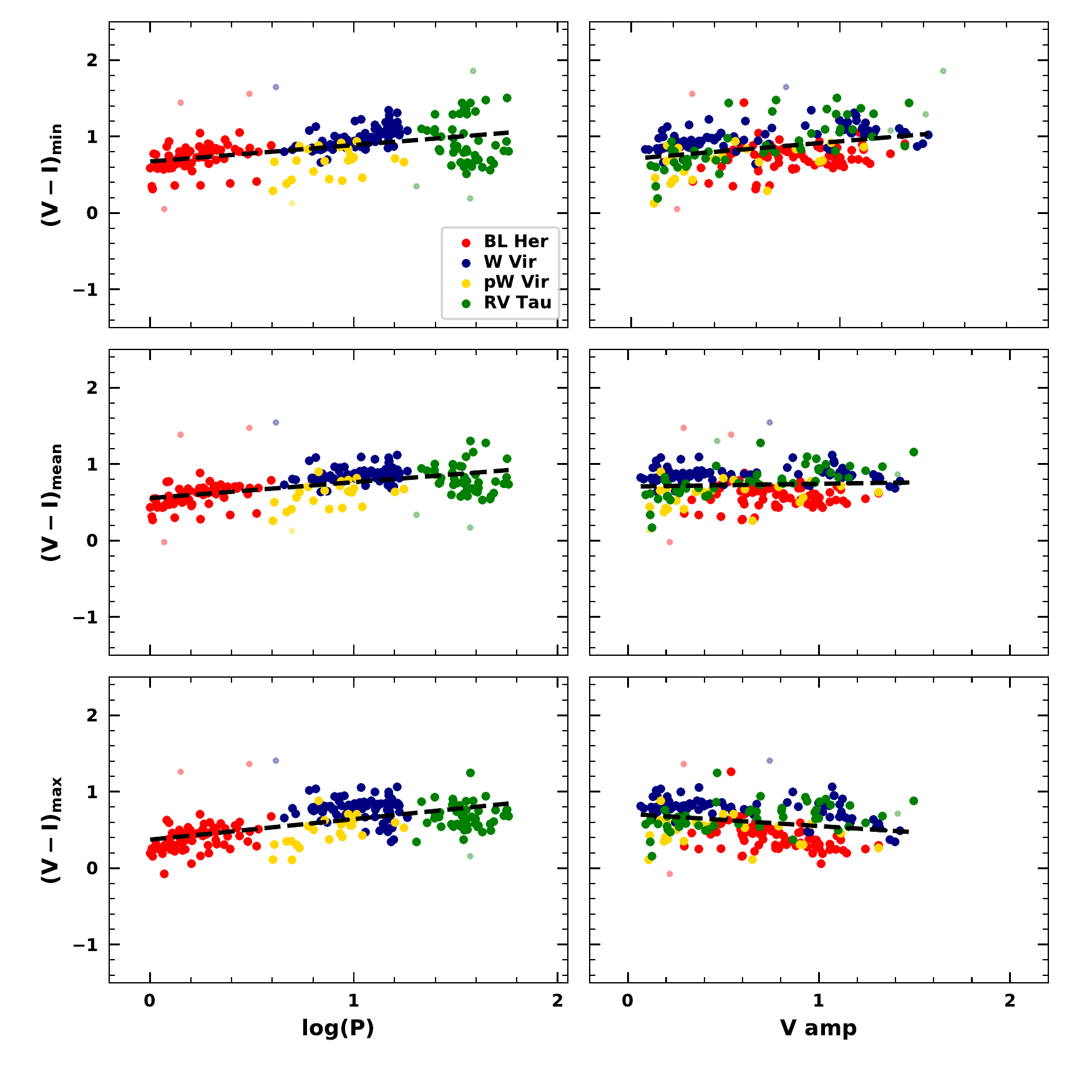}
\caption{Same as Fig.\ref{fig:T2C_Bulge} but for LMC.}
\label{fig:T2C_LMC} 
\end{figure*}

\begin{figure*}
\centering
\includegraphics[scale = 0.85]{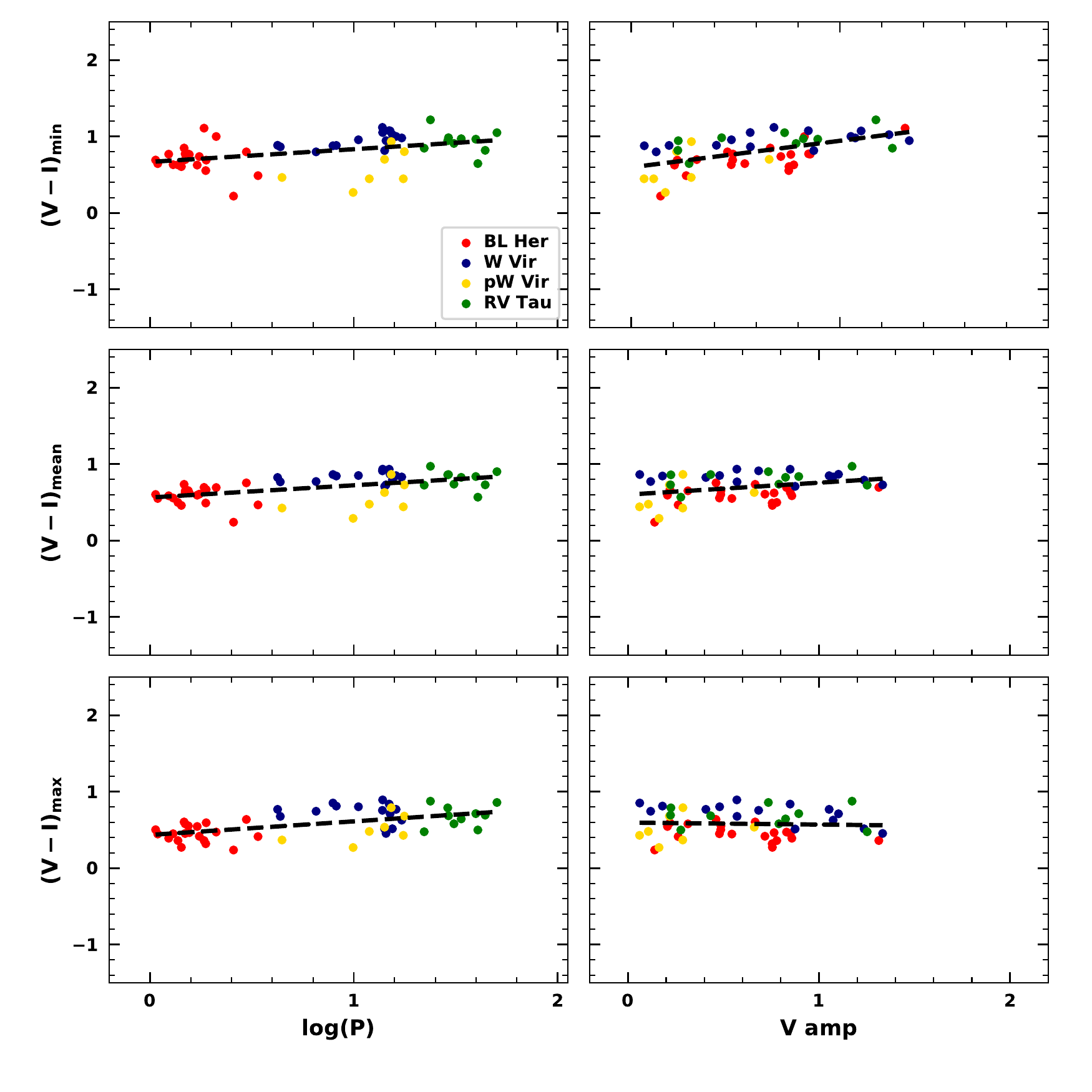}
\caption{Same as Fig.\ref{fig:T2C_Bulge} but for SMC.}
\label{fig:T2C_SMC}
\end{figure*}

\begin{figure*}
\centering
\includegraphics[scale = 0.85]{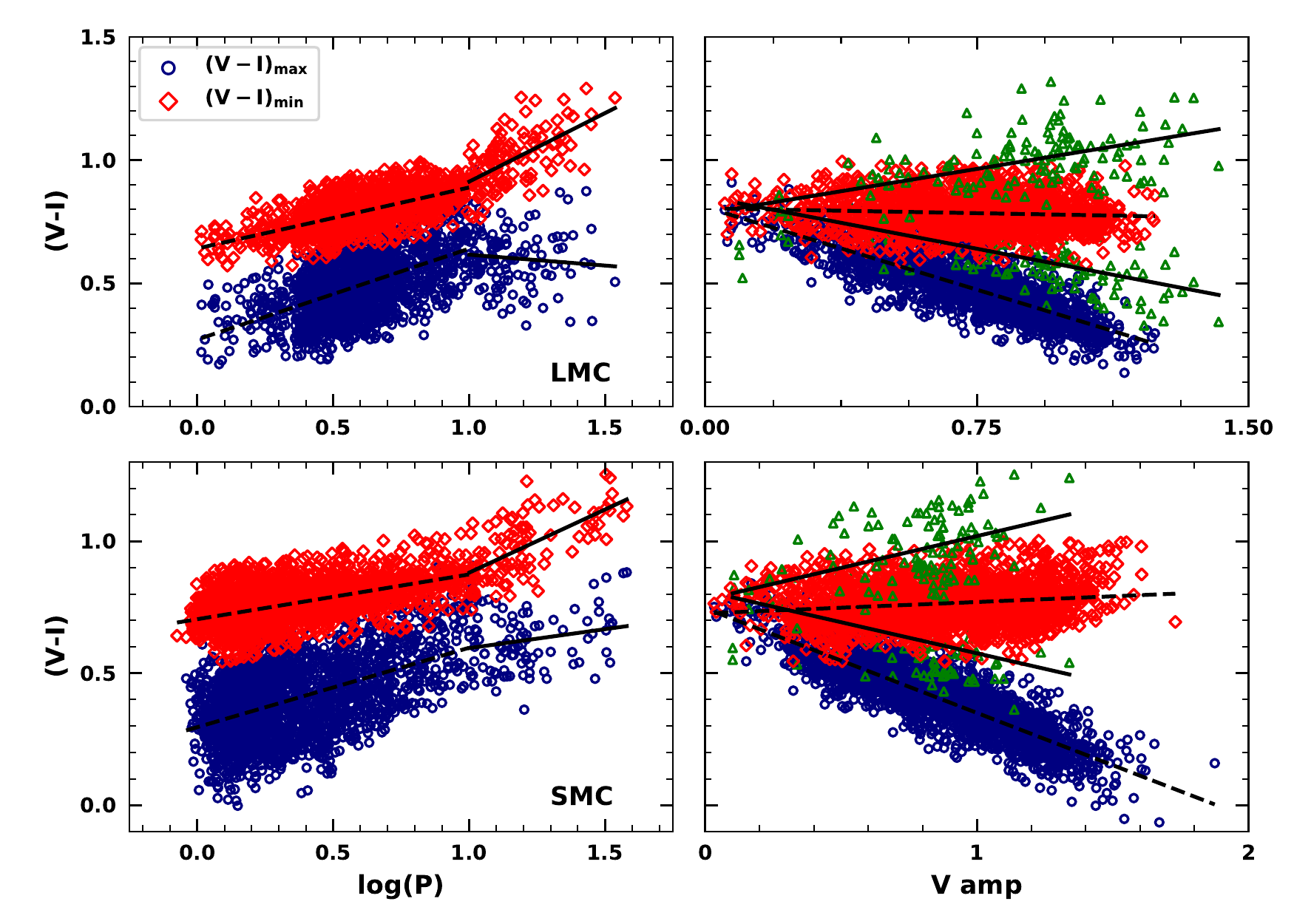}
\caption{PC and AC relations for fundamental mode classical Cepheids from OGLE-IV at maximum and minimum light for LMC and SMC. The dashed/solid lines represent the best fit to shorter/longer period Cepheids separated at 10 days. Green triangles on the right represent the Cepheids having periods greater than 10 days at both maximum and minimum light.}
\label{fig:PCAC_Cep}
\end{figure*}

\section{Analysis and Results}
\label{sec:PCAC}
\subsection{The statistical tests}
\subsubsection{The $F$-test}

Earlier studies by \citet{kanbur2004a} and \citet{bhardwaj2014} showed a break in the PC relations for fundamental mode classical Cepheids in LMC and SMC at a period of 10 days at both minimum and maximum light. To test for any possible break in the PC relations for OGLE-IV classical Cepheids in the Magellanic Cloud, we use the $F$-test, which has been well-described in \citet{kanbur2004a} and \citet{bhardwaj2014}. We briefly outline the method here: a single regression line may fit the data over the entire period range under the null hypothesis while two separate regression lines are required to fit the data having periods shorter/longer than 10 days under the alternate hypothesis, where we are testing for a break in the PC relation at a period of 10 days. The reduced model under the null hypothesis may then be written as:
\begin{equation}
(V-I)_\textrm{{max/min}}=a+b\log(P),
\end{equation}
and the full model under the alternate hypothesis is:
\begin{equation}
\begin{aligned}
(V-I)_\textrm{{max/min}}&=a_S+b_S\log(P); \textrm{where $P<$10 days}\\
&=a_L+b_L\log(P); \textrm{where $P\geq$10 days},
\end{aligned}
\end{equation}
where $(V-I)_\textrm{{max/min}}$ is colour at maximum or minimum light.

The $F$ statistic is defined as follows \citep[see example,][]{kanbur2004a,bhardwaj2014}:

\begin{equation}
F = \frac{RSS_R-RSS_F/[(n-2)-(n-4)]}{RSS_F/(n-4)},
\end{equation}
where $RSS_R$, $RSS_F$ are the residual sums of squares in the reduced model and the full model, respectively and $n$ is the total number of stars in the entire dataset. The statistical significance of the variation in slopes at periods shorter/longer than 10 days may be studied by comparing the observed value of the $F$ statistic with the critical value, $F_C=F_{2,n-4}$ at 95\% confidence level. An $F$ statistic greater than the critical value or the probability of the observed value of the $F$ statistic, P(F) $<$ 0.05 would mean the rejection of the null hypothesis, thereby confirming the presence of a break in the PC relation at 10 days.

\subsubsection{The $t$-test}

We use the standard $t$-test to statistically check the equivalence of PC slopes of different types of pulsating variable stars. We briefly outline the method here; detailed description may be found in \citet{ngeow2015}. The $T$ statistic to compare slopes, $\hat{W}$ of two linear regressions with sample sizes, $n$ and $m$, respectively is defined as:
\begin{equation}
T=\frac{\hat{W}_n-\hat{W}_m}{\sqrt{\mathrm{Var}(\hat{W}_n)+\mathrm{Var}(\hat{W}_m)}},
\end{equation}
where $\mathrm{Var}(\hat{W})$ is the variance of the slope. The critical value under the two-tailed $t$-distribution is $t_{\alpha/2,\nu}$, where $\alpha$=0.05 with 95\% confidence limit and degrees of freedom, $\nu=n+m-4$. The null hypothesis of equivalent PC slopes may be rejected if $T>t_{\alpha/2,\nu}$ or the probability of the observed value of the $T$ statistic, $p<0.05$.

\begin{table*}
\caption{The slopes and intercepts for PC and AC relations for the four subclasses of the T2Cs in the Bulge, LMC and SMC. T2C refers to these subclasses combined together. The last column indicates whether the PC relation has a flat or a significant slope within 3$\sigma$ uncertainties.}
\centering
\scalebox{0.9}{
\begin{tabular}{c c c c c c c c c}
\hline\hline
& Phase & \multicolumn{3}{c}{PC} & \multicolumn{3}{c}{AC} & Nature of PC slope\\
\hline \hline
& & Slope & Intercept & $\sigma$ & Slope & Intercept & $\sigma$ &\\
\hline
\multicolumn{9}{c}{Bulge (OGLE-IV) using extinction law from \citet{cardelli1989}}\\
\hline
BL~Her & min & 0.516$\pm$0.073 & 0.475$\pm$0.023 & 0.173 & 0.216$\pm$0.053 & 0.45$\pm$0.045 & 0.219 & Sloped\\
& mean & 0.594$\pm$0.072 & 0.323$\pm$0.022 & 0.17 & -0.057$\pm$0.05 & 0.518$\pm$0.043 & 0.208 & Sloped\\
& max & 0.636$\pm$0.088 & 0.122$\pm$0.028 & 0.21 & -0.38$\pm$0.052 & 0.584$\pm$0.044 & 0.216 & Sloped\\
W~Vir & min & 0.549$\pm$0.136 & 0.289$\pm$0.141 & 0.257 & 0.193$\pm$0.048 & 0.725$\pm$0.036 & 0.253 & Sloped\\
& mean & 0.276$\pm$0.122 & 0.444$\pm$0.126 & 0.228 & 0.032$\pm$0.047 & 0.72$\pm$0.036 & 0.247 & Flat\\
& max & 0.037$\pm$0.141 & 0.605$\pm$0.147 & 0.266 & -0.172$\pm$0.05 & 0.751$\pm$0.038 & 0.266 & Flat\\
pW~Vir & min & -0.819$\pm$0.665 & 1.659$\pm$0.633 & 0.15 & 0.295$\pm$0.186 & 0.676$\pm$0.136 & 0.145 & Flat\\
& mean & -0.487$\pm$0.671 & 1.232$\pm$0.639 & 0.151 & 0.089$\pm$0.196 & 0.707$\pm$0.144 & 0.153 & Flat\\
& max & -0.23$\pm$0.769 & 0.85$\pm$0.732 & 0.173 & -0.191$\pm$0.217 & 0.765$\pm$0.159 & 0.169 & Flat\\
RV Tau & min & 0.03$\pm$0.317 & 1.037$\pm$0.475 & 0.359 & 0.207$\pm$0.103 & 0.877$\pm$0.109 & 0.35 & Flat\\
& mean & 0.24$\pm$0.298 & 0.532$\pm$0.446 & 0.337 & -0.084$\pm$0.099 & 0.972$\pm$0.105 & 0.337 & Flat\\
& max & 0.47$\pm$0.333 & 0.053$\pm$0.499 & 0.379 & -0.322$\pm$0.103 & 1.071$\pm$0.108 & 0.334 & Flat\\
T2C & min & 0.35$\pm$0.023 & 0.522$\pm$0.021 & 0.244 & 0.196$\pm$0.038 & 0.651$\pm$0.032 & 0.307 & Sloped\\
& mean & 0.327$\pm$0.02 & 0.389$\pm$0.019 & 0.217 & -0.014$\pm$0.036 & 0.668$\pm$0.031 & 0.293 & Sloped\\
& max & 0.396$\pm$0.024 & 0.202$\pm$0.022 & 0.26 & -0.236$\pm$0.041 & 0.697$\pm$0.035 & 0.335 & Sloped\\
\hline
\multicolumn{9}{c}{Bulge (OGLE-IV) using extinction law from \citet{nataf2013}}\\
\hline
BL~Her & min & 0.389$\pm$0.058 & 0.58$\pm$0.018 & 0.107 & 0.063$\pm$0.042 & 0.639$\pm$0.034 & 0.13 & Sloped\\
& mean & 0.337$\pm$0.066 & 0.472$\pm$0.021 & 0.123 & -0.097$\pm$0.04 & 0.63$\pm$0.033 & 0.123 & Sloped\\
& max & 0.374$\pm$0.102 & 0.298$\pm$0.032 & 0.194 & -0.376$\pm$0.043 & 0.667$\pm$0.036 & 0.134 & Sloped\\
W~Vir & min & 0.383$\pm$0.116 & 0.482$\pm$0.119 & 0.161 & 0.183$\pm$0.042 & 0.764$\pm$0.03 & 0.156 & Sloped\\
& mean & 0.043$\pm$0.108 & 0.736$\pm$0.112 & 0.153 & 0.021$\pm$0.041 & 0.768$\pm$0.029 & 0.153 & Flat\\
& max & -0.151$\pm$0.12 & 0.848$\pm$0.124 & 0.169 & -0.129$\pm$0.044 & 0.77$\pm$0.031 & 0.164 & Flat\\
pW~Vir & min & -0.37$\pm$0.765 & 1.263$\pm$0.733 & 0.139 & 0.313$\pm$0.19 & 0.709$\pm$0.129 & 0.122 & Flat\\
& mean & -0.148$\pm$0.784 & 0.956$\pm$0.751 & 0.143 & 0.156$\pm$0.216 & 0.715$\pm$0.146 & 0.139 & Flat\\
& max & 0.121$\pm$0.894 & 0.581$\pm$0.857 & 0.163 & -0.104$\pm$0.251 & 0.763$\pm$0.17 & 0.162 & Flat\\
RV Tau & min & 0.111$\pm$0.319 & 1.013$\pm$0.478 & 0.298 & 0.154$\pm$0.112 & 1.034$\pm$0.114 & 0.293 & Flat\\
& mean & 0.419$\pm$0.298 & 0.373$\pm$0.447 & 0.279 & -0.136$\pm$0.107 & 1.126$\pm$0.108 & 0.28 & Flat\\
& max & 0.766$\pm$0.349 & -0.267$\pm$0.524 & 0.33 & -0.404$\pm$0.101 & 1.262$\pm$0.102 & 0.262 & Flat\\
T2C & min & 0.321$\pm$0.019 & 0.594$\pm$0.017 & 0.16 & 0.206$\pm$0.038 & 0.69$\pm$0.031 & 0.231 & Sloped\\
& mean & 0.316$\pm$0.017 & 0.472$\pm$0.015 & 0.14 & 0.031$\pm$0.037 & 0.697$\pm$0.03 & 0.227 & Sloped\\
& max & 0.36$\pm$0.023 & 0.31$\pm$0.021 & 0.194 & -0.2$\pm$0.042 & 0.728$\pm$0.035 & 0.261 & Sloped\\
\hline
\multicolumn{9}{c}{LMC (OGLE-IV)}\\
\hline
BL~Her & min & 0.604$\pm$0.101 & 0.602$\pm$0.026 & 0.112 & 0.195$\pm$0.079 & 0.566$\pm$0.064 & 0.146 & Sloped\\
& mean & 0.585$\pm$0.062 & 0.453$\pm$0.016 & 0.067 & 0.009$\pm$0.069 & 0.571$\pm$0.056 & 0.127 & Sloped\\
& max & 0.44$\pm$0.104 & 0.293$\pm$0.027 & 0.121 & -0.263$\pm$0.066 & 0.597$\pm$0.054 & 0.122 & Sloped\\
W~Vir & min & 0.597$\pm$0.072 & 0.366$\pm$0.074 & 0.102 & 0.193$\pm$0.029 & 0.866$\pm$0.02 & 0.11 & Sloped\\
& mean & 0.203$\pm$0.059 & 0.651$\pm$0.061 & 0.082 & 0.018$\pm$0.023 & 0.85$\pm$0.016 & 0.09 & Sloped\\
& max & -0.033$\pm$0.086 & 0.821$\pm$0.088 & 0.12 & -0.137$\pm$0.032 & 0.855$\pm$0.022 & 0.122 & Flat\\
pW~Vir & min & 0.468$\pm$0.26 & 0.235$\pm$0.233 & 0.195 & 0.322$\pm$0.118 & 0.49$\pm$0.07 & 0.18 & Flat\\
& mean & 0.528$\pm$0.209 & 0.116$\pm$0.187 & 0.157 & 0.157$\pm$0.113 & 0.504$\pm$0.067 & 0.172 & Flat\\
& max & 0.647$\pm$0.195 & -0.091$\pm$0.174 & 0.146 & -0.113$\pm$0.117 & 0.533$\pm$0.07 & 0.178 & Sloped\\
RV Tau & min & -0.082$\pm$0.461 & 1.062$\pm$0.715 & 0.33 & 0.655$\pm$0.077 & 0.519$\pm$0.058 & 0.206 & Flat\\
& mean & -0.088$\pm$0.291 & 0.918$\pm$0.451 & 0.208 & 0.32$\pm$0.04 & 0.566$\pm$0.031 & 0.105 & Flat\\
& max & 0.003$\pm$0.198 & 0.661$\pm$0.308 & 0.141 & 0.178$\pm$0.053 & 0.542$\pm$0.04 & 0.141 & Flat\\
T2C & min & 0.215$\pm$0.026 & 0.673$\pm$0.026 & 0.195 & 0.229$\pm$0.04 & 0.708$\pm$0.029 & 0.221 & Sloped\\
& mean & 0.209$\pm$0.02 & 0.555$\pm$0.02 & 0.151 & 0.037$\pm$0.033 & 0.706$\pm$0.024 & 0.185 & Sloped\\
& max & 0.269$\pm$0.024 & 0.372$\pm$0.024 & 0.182 & -0.161$\pm$0.039 & 0.711$\pm$0.029 & 0.22 & Sloped\\
\hline
\multicolumn{9}{c}{SMC (OGLE-IV)}\\
\hline
BL~Her & min & -0.251$\pm$0.339 & 0.759$\pm$0.088 & 0.175 & 0.434$\pm$0.115 & 0.445$\pm$0.076 & 0.129 & Flat\\
& mean & -0.115$\pm$0.22 & 0.613$\pm$0.057 & 0.114 & 0.127$\pm$0.097 & 0.512$\pm$0.064 & 0.109 & Flat\\
& max & -0.018$\pm$0.21 & 0.458$\pm$0.054 & 0.108 & -0.12$\pm$0.092 & 0.525$\pm$0.06 & 0.103 & Flat\\
W~Vir & min & 0.314$\pm$0.112 & 0.632$\pm$0.118 & 0.074 & 0.124$\pm$0.063 & 0.869$\pm$0.051 & 0.083 & Flat\\
& mean & 0.073$\pm$0.098 & 0.757$\pm$0.103 & 0.064 & -0.032$\pm$0.049 & 0.854$\pm$0.04 & 0.065 & Flat\\
& max & -0.169$\pm$0.189 & 0.89$\pm$0.199 & 0.125 & -0.219$\pm$0.073 & 0.87$\pm$0.059 & 0.096 & Flat\\
pW~Vir & min & 0.52$\pm$0.576 & 0.02$\pm$0.631 & 0.194 & 0.52$\pm$0.618 & 0.448$\pm$0.194 & 0.197 & Flat\\
& mean & 0.478$\pm$0.473 & 0.035$\pm$0.518 & 0.159 & 0.37$\pm$0.538 & 0.456$\pm$0.169 & 0.171 & Flat\\
& max & 0.484$\pm$0.408 & -0.014$\pm$0.447 & 0.137 & 0.216$\pm$0.508 & 0.453$\pm$0.16 & 0.162 & Flat\\
RV Tau & min & -0.366$\pm$0.507 & 1.492$\pm$0.773 & 0.137 & 0.199$\pm$0.14 & 0.799$\pm$0.108 & 0.123 & Flat\\
& mean & -0.203$\pm$0.394 & 1.11$\pm$0.601 & 0.106 & 0.105$\pm$0.116 & 0.73$\pm$0.089 & 0.102 & Flat\\
& max & 0.198$\pm$0.475 & 0.38$\pm$0.725 & 0.128 & -0.003$\pm$0.148 & 0.683$\pm$0.114 & 0.13 & Flat\\
T2C & min & 0.166$\pm$0.054 & 0.668$\pm$0.052 & 0.196 & 0.347$\pm$0.072 & 0.597$\pm$0.05 & 0.176 & Sloped\\
& mean & 0.16$\pm$0.04 & 0.563$\pm$0.039 & 0.145 & 0.156$\pm$0.064 & 0.601$\pm$0.045 & 0.158 & Sloped\\
& max & 0.177$\pm$0.04 & 0.435$\pm$0.039 & 0.147 & -0.026$\pm$0.071 & 0.596$\pm$0.049 & 0.174 & Sloped\\
\hline
\end{tabular}
}
\label{tab:PCAC}
\end{table*}

\subsection{Period-colour and amplitude-colour relations} 
\citet{simon1993} introduced the concept of PCAC at maximum and minimum light as a way to probe the radiation hydrodynamics of the outer envelope in Cepheids and were able to explain the flat and sloped PC relations for Galactic Cepheids at maximum and minimum light, respectively. This was extended in a series of papers \citep[and references therein]{bhardwaj2014, das2018}. Here we study PCAC relations for T2Cs and classical Cepheids using the latest available data.

\subsubsection{Type II Cepheids}
The period-colour (PC) and amplitude-colour (AC) relations for the T2Cs in the Bulge, LMC and SMC from OGLE-IV are displayed in Figs.~\ref{fig:T2C_Bulge}, ~\ref{fig:T2C_LMC} and ~\ref{fig:T2C_SMC}, respectively at minimum, mean and maximum light. PC and AC relations are also obtained for different sub-classes of T2Cs separately. The summary of the PC and AC relations for different sub-classes as well as for all the T2Cs sample in the Bulge, LMC and SMC are listed in Table~\ref{tab:PCAC}. First we discuss the PC relations of the W~Vir stars. We observe a flat PC slope at maximum light for W~Vir stars in the Bulge, LMC and SMC within 3$\sigma$ uncertainties. For W~Vir stars, the PC$_\textrm{max}$ is flat but slightly negative ($-0.033 \pm 0.086$ in the LMC and $-0.169 \pm 0.189$ in the SMC). The flat PC$_\textrm{max}$ slope of W~Vir stars is similar to that obtained for long period classical Cepheids ($P>10 \textrm{days}$) in Galaxy, LMC and SMC \citep{bhardwaj2014}. We note that the slope of PC relation at minimum light for W~Vir stars is significantly greater than at maximum light. In fundamental mode classical Cepheids, the HIF and stellar photosphere are engaged at low densities only at maximum light leading to a flat PC relation. Apparent similarity in the behaviour of W Vir stars with long-period classical Cepheids in the PCAC plane suggests that HIF and stellar photosphere interactions are similar in these two classical pulsators as they also occupy the same region on the HR diagram.

In the PC relations of BL~Her stars, we observe a significant, positive PC slope in the Bulge and the LMC and a flat (but negative) slope in the SMC at both minimum and maximum light. The sloped PC relations of BL Her stars  in the Bulge and the LMC are similar to short period classical Cepheids ($P<10 \textrm{days}$). The flat but slightly negative PC$_\textrm{min}$ for BL~Her stars in SMC ($-0.251 \pm 0.339$) is similar to what \citet{das2018} obtained for RR~Lyrae stars in SMC at minimum light. We note here that the number of BL~Her stars in SMC is very less ($\sim 20$). The slopes of the PC relations for RV~Tau stars in Bulge, LMC and SMC are flat within 3$\sigma$ uncertainties at minimum and maximum light; however, there is high dispersion in all the cases.

\subsubsection{Classical Cepheids}
The PC and AC relations for classical Cepheids in LMC and SMC from OGLE-IV are shown in Fig.~\ref{fig:PCAC_Cep} and summarised in Table~\ref{tab:PCAC_cep}. We have used the $F$-test to statistically identify any possible breaks in the PC and AC relations. As can be observed from Table~\ref{tab:PCAC_cep}, we find breaks in the PC as well as AC relations at a period of 10 days at both minimum and maximum light for classical Cepheids in LMC and SMC. The longer period classical Cepheids have a flat PC$_{\mathrm{max}}$ within 3$\sigma$ uncertainties and a significantly sloped PC$_{\mathrm{min}}$ relation for both LMC and SMC. Also, at maximum light, longer period classical Cepheids have a shallower PC$_{\mathrm{max}}$ than their shorter period counterparts. This is similar to the results found in the earlier studies for classical Cepheids in LMC and SMC by \citet{kanbur2004a} and \citet{bhardwaj2014}. We have excluded classical Cepheids in the Galactic bulge from any statistical test because of the availability of very few stars from the OGLE-IV dataset.
We note that the PC relations at maximum/minimum show statistically different slopes between long and short period Cepheids and between the LMC and the SMC. In future work, these differences can be used to constrain extensive grids of Cepheid pulsation models.

\begin{table*}
\caption{Results of F test on PC and AC relations for fundamental mode Cepheids from OGLE-IV to determine possible nonlinearities at 10 days. The bold-face entries indicate existence of break in PC/AC at 10 days.}
\centering
\begin{tabular}{c c c c c c c c c c}
\hline\hline
& Phase & \textrm{$b_{all}$} & \textrm{$a_{all}$} & \textrm{$b_{S}$} & \textrm{$a_{S}$} & \textrm{$b_{L}$} & \textrm{$a_{L}$} & F & P(F)\\
\hline \hline
\multicolumn{10}{c}{LMC}\\
\hline
PC & max & 0.294$\pm$0.011 & 0.312$\pm$0.007 & 0.369$\pm$0.015 & 0.272$\pm$0.009 & -0.088$\pm$0.087 & 0.705$\pm$0.101 & 28.448 & \bf{0.000}\\
& mean & 0.246$\pm$0.006 & 0.539$\pm$0.004 & 0.228$\pm$0.008 & 0.549$\pm$0.005 & 0.419$\pm$0.062 & 0.358$\pm$0.072 & 2.772 & 0.063\\
& min & 0.294$\pm$0.006 & 0.618$\pm$0.004 & 0.247$\pm$0.008 & 0.642$\pm$0.004 & 0.556$\pm$0.067 & 0.356$\pm$0.078 & 11.202 & \bf{0.000}\\
AC & max & -0.421$\pm$0.008 & 0.797$\pm$0.006 & -0.447$\pm$0.007 & 0.81$\pm$0.006 & -0.281$\pm$0.032 & 0.851$\pm$0.03 & 143.275 & \bf{0.000}\\
& mean & -0.132$\pm$0.007 & 0.784$\pm$0.006 & -0.148$\pm$0.007 & 0.791$\pm$0.005 & 0.109$\pm$0.032 & 0.743$\pm$0.029 & 145.555 & \bf{0.000}\\
& min & -0.002$\pm$0.008 & 0.793$\pm$0.007 & -0.026$\pm$0.008 & 0.804$\pm$0.006 & 0.242$\pm$0.036 & 0.783$\pm$0.033 & 180.097 & \bf{0.000}\\
\hline
\multicolumn{10}{c}{SMC}\\
\hline
PC & max & 0.284$\pm$0.009 & 0.301$\pm$0.004 & 0.301$\pm$0.011 & 0.296$\pm$0.005 & 0.143$\pm$0.07 & 0.452$\pm$0.085 & 7.141 & \bf{0.001}\\
& mean & 0.205$\pm$0.004 & 0.569$\pm$0.002 & 0.188$\pm$0.005 & 0.573$\pm$0.002 & 0.432$\pm$0.055 & 0.316$\pm$0.066 & 5.093 & \bf{0.006}\\
& min & 0.197$\pm$0.005 & 0.696$\pm$0.002 & 0.17$\pm$0.006 & 0.705$\pm$0.002 & 0.478$\pm$0.056 & 0.402$\pm$0.068 & 13.24 & \bf{0.000}\\
AC & max & -0.398$\pm$0.005 & 0.751$\pm$0.005 & -0.396$\pm$0.005 & 0.746$\pm$0.005 & -0.236$\pm$0.044 & 0.811$\pm$0.036 & 218.073 & \bf{0.000}\\
& mean & -0.113$\pm$0.004 & 0.74$\pm$0.004 & -0.111$\pm$0.004 & 0.736$\pm$0.004 & 0.082$\pm$0.049 & 0.768$\pm$0.04 & 227.841 & \bf{0.000}\\
& min & 0.042$\pm$0.006 & 0.731$\pm$0.005 & 0.043$\pm$0.005 & 0.726$\pm$0.005 & 0.241$\pm$0.05 & 0.778$\pm$0.041 & 193.507 & \bf{0.000}\\
\hline
\end{tabular}
\label{tab:PCAC_cep}
\end{table*}

\subsubsection{Comparison of Period-Colour relations}
\label{sec:t-test}
It is important to quantitatively compare the PC relations at maximum, mean and minimum light between the different classes of variable stars and different
metallicity environments. Since previous work \citep[and references therein]{bhardwaj2014} has suggested nonlinearities in PC relations, we investigate whether the data used in this
paper show evidence of this. Nonlinearities in PC relations will be useful in constraining stellar pulsation and evolution models.

We summarize the results of the statistical $t$-test to compare PC slopes in the LMC and SMC across a broad spectrum of variable star types in Tables~\ref{tab:ttest_LMC} and \ref{tab:ttest_SMC}, respectively. The PC slopes of RRab stars in LMC and SMC using OGLE-IV data have been taken from our recent work \citep{das2018}. We discuss the results from this comparison below:-
\begin{enumerate}

\item Slopes of the PC relation for W~Vir stars are equivalent with those for the longer-period classical Cepheids at both minimum and maximum light for LMC but only at 
minimum light for SMC. 

\item The PC$_{\mathrm{min}}$ slopes for BL~Her stars and RRab stars in SMC are statistically similar.

\item BL~Her stars have statistically similar PC slopes with those from longer period classical Cepheids in LMC at minimum light and with those in SMC at maximum light.

\item In SMC, BL~Her stars have PC slopes equivalent with those from W~Vir stars at both minimum and maximum light while in LMC, they are statistically similar only at minimum light. 

\item BL~Her stars have PC slopes equivalent with those from shorter period classical Cepheids at both minimum and maximum light in SMC but only at maximum light for those in LMC.

\item For both LMC and SMC, PC slopes from W~Vir stars and RRab stars exhibit different behaviour at minimum and maximum light. 

\item As expected, RRab stars and classical Cepheids never have similar PC slopes; this is because of the existence of flat PC$_{\mathrm{min}}$ for RRab with a significantly sloped PC$_{\mathrm{max}}$ and flat PC$_{\mathrm{max}}$ for classical Cepheids with a significant slope in PC$_{\mathrm{min}}$.

\end{enumerate}

The central result from these tables is the following: RRab stars have a flat or shallow slope at minimum light and a statistically significant larger slope at maximum light. BL Her stars have a much smaller difference between the PC slopes at maximum/minimum light than RRab stars, similar to short-period classical Cepheids. WVir stars, in contrast, have statistically significant flat or shallow (perhaps negative) slopes at maximum light and a statistically significant non-zero slope at minimum light - a situation that is similar to that of long-period classical Cepheids. 
There may be further real differences between different galaxies, but for this paper we concentrate on this broad conclusion and seek to explain this in terms of the detailed physics of the outer envelope and how this changes as a result of stellar evolution.

We also see that the PCAC analysis of OGLE-IV data displayed in Fig.~\ref{fig:PCAC_Cep} and quantified in Table~\ref{tab:PCAC_cep} portray very clearly a sharp break in the PCAC relations at a period close to 10 days. This is consistent with previous work by \citet{bhardwaj2014} using OGLE-III data. The interpretation of this break in the PC relation arises from arguments given in \citet{kanbur2007,bhardwaj2014}, and references therein regarding the interaction of the HIF and stellar photosphere. For periods shorter than about 10 days, the HIF and stellar photosphere are engaged at high temperatures and hence the temperature of the photosphere is the temperature at which hydrogen ionizes. This is in a regime where Saha ionization equilibrium is sensitive to temperature and density and hence, sensitive to global stellar parameters. Since colour is a measure of the temperature of the photosphere and period is dependent on global stellar parameters through the period-mean density theorem, we would expect a non-zero slope for the PC relation. However, as the $L/M$ ratio goes up and the temperature goes down, the period increases and the HIF and photosphere become disengaged at many phases, except around maximum light when the HIF is at its furthest point in the mass distribution. This engagement at maximum light at lower temperatures and densities implies that the temperature of the stellar photosphere is again the temperature at which hydrogen ionizes but this time the temperature at which hydrogen ionizes is somewhat independent of global stellar parameters and hence we have a flatter PC slope for these long period Cepheids. The AC part of PCAC arise from arguments first presented in \citet{simon1993}: changes in PC relations will be reflected in AC relations and this is what we find. OGLE-IV PCAC results for RR Lyrae stars are presented in \citet{das2018} and again are consistent with the above ideas. Hence it is important to see how the latest available data for T2Cs in different metallicity environments compare to results for Cepheids and RR Lyraes.

\section{HIF-Stellar Photosphere interaction - The theoretical approach}
\label{sec:HIF-Ph}

\begin{table*}
\caption{A summary of the RR~Lyrae, BL~Her and classical Cepheid models computed with MESA for the present analysis with a unique combination of ($Z$, $X$, $M/M_{\odot}$, $L/L_{\odot}$ and ${T_\text{eff}}$)). $\mathrm{P_A}$ and $\mathrm{P_D}$ indicate period in days using convection sets A and D, respectively. All the models presented here are fundamental mode pulsators. The entries marked with ``-'' are those that did not have full-amplitude stable pulsations in the fundamental mode.}
\label{tab:Data}
\centering
\scalebox{0.9}{
\begin{tabular}{c c c c c c c}
\hline
\hline
Z & X & $\dfrac{M}{M_{\odot}}$ & $\dfrac{L}{L_{\odot}}$ & ${T_\text{eff}}$ (K) & $\mathrm{P_A}$ & $\mathrm{P_D}$
\\ [0.5ex]
\hline
\hline
\multicolumn{7}{c}{RR Lyrae models}\\
\hline
0.001 & 0.754 & 0.64 & 46.77 & 6800 &0.49833&0.36771\\
0.001 & 0.754 & 0.64 & 46.77 & 6300 &0.65492&0.65561\\
0.001 & 0.754 & 0.58 & 74.13 & 6900 &0.73759&$-$\\
0.001 & 0.754 & 0.58 & 74.13 & 6700 &0.82012&0.60037\\
0.001 & 0.754 & 0.64 & 97.72 & 6600 &1.02266&1.02203\\
0.001 & 0.754 & 0.64 & 97.72 & 6700 &0.96767&$-$\\
0.02 & 0.71 & 0.54 & 30.90 & 6800 &0.41861&0.29709\\
0.02 & 0.71 & 0.54 & 30.90 & 6700 &0.44036&0.31198\\
0.02 & 0.71 & 0.54 & 30.90 & 6500 &0.48886&0.48887\\
0.004 & 0.746 & 0.59 & 40.74 & 6900 &0.45505&0.33229\\
0.0006 & 0.7544 & 0.67 & 48.98 & 6800 &0.49924&0.37066\\
0.008 & 0.736 & 0.56 & 39.81 & 6900 &0.46738&0.33825\\
0.001 & 0.754 & 0.64 & 46.77 & 6200 &0.69655&0.69808\\
0.001 & 0.754 & 0.64 & 46.77 & 6400 &0.6178&0.61794\\
0.001 & 0.754 & 0.64 & 46.77 & 6500 &0.58415&0.58414\\
0.001 & 0.754 & 0.64 & 46.77 & 6600 &0.55327&0.55312\\
0.001 & 0.754 & 0.64 & 46.77 & 6700 &0.52477&0.36918\\
\hline
\multicolumn{7}{c}{BL~Her models}\\
\hline
0.0001 & 0.7599 & 0.6 & 89.12 & 6300 &1.16531&1.1663\\
0.001 & 0.759 & 0.55 & 64.56 & 6250 &0.98055&0.98211\\
0.001 & 0.759 & 0.65 & 81.28 & 6200 &1.09424&1.09615\\
0.001 & 0.759 & 0.65 & 64.56 & 6200 &0.90085&0.90257\\
0.001 & 0.759 & 0.65 & 81.28 & 6600 &0.8668&0.63547\\
0.001 & 0.759 & 0.65 & 102.33 & 6200 &1.33299&1.33625\\
0.004 & 0.756 & 0.55 & 102.33 & 5950 &1.81067&1.82324\\
0.0001 & 0.7599 & 0.65 & 128.82 & 5750 &2.20416&2.25865\\
0.0001 & 0.7599 & 0.6 & 141.25 & 5750 &2.52525&2.58125\\
0.0004 & 0.75 & 0.6 & 150 & 5780 &2.60992&2.65789\\
0.0004 & 0.75 & 0.6 & 175 & 5780 &2.98105&3.03453\\
0.0004 & 0.75 & 0.6 & 300 & 5780 &4.81726&4.88523\\
0.004 & 0.756 & 0.55 & 64.56 & 5950 &1.21133&1.22078\\
0.004 & 0.756 & 0.55 & 64.56 & 6750 &0.75042&0.54628\\
0.004 & 0.756 & 0.55 & 81.28 & 5850 &$1.61118^*$&1.61118\\
0.004 & 0.756 & 0.55 & 81.28 & 6650 &0.9599&0.95901\\
0.004 & 0.756 & 0.55 & 102.33 & 5650 &2.24159&2.31228\\
0.0001 & 0.7599 & 0.65 & 128.82 & 6225 &1.58928&1.589\\
0.0001 & 0.7599 & 0.6 & 141.25 & 5600 &2.85031&2.99971\\
\hline
\multicolumn{7}{c}{Classical Cepheid models}\\
\hline
0.02 & 0.7 & 5.1 & 3006.08 & 5396 &10.32019&10.32818\\
0.02 & 0.7 & 7.7 & 4965.92 & 5332 &12.46112&$-$\\
0.02 & 0.7 & 6.3 & 6456.54 & 5390 &17.69231&17.68952\\
0.02 & 0.7 & 9.45 & 9862.79 & 5265 &20.88859&$-$\\
0.004 & 0.746 & 6.8 & 8709.64 & 5200 &24.49454&24.84936\\
0.004 & 0.746 & 6.8 & 8709.64 & 5000 &28.59717&29.59406\\
0.004 & 0.746 & 6.6 & 7762.47 & 4900 &28.51875&30.12038\\
0.004 & 0.746 & 6.6 & 8709.64 & 4900 &31.74634&33.49494\\	
0.004 & 0.746 & 6.8 & 8709.64 & 4800 &$-$&36.41581\\
0.02 & 0.7 & 10 & 34197.94 & 5100 &78.4023&$-$\\
0.02 & 0.7 & 9 & 23388.37 & 5100 &57.05626&58.18299\\
0.008 & 0.736 & 4.5 & 1258.92 & 5960 &3.59225&2.55345\\
0.008 & 0.736 & 6 & 3981.07 & 5625 &9.8025&9.79768\\
0.008 & 0.736 & 8 & 10000 & 5370 &21.88827&21.9776\\
0.008 & 0.736 & 10 & 19952.62 & 5310 &36.90233&37.16359\\
0.008 & 0.742 & 5.6 & 1995.26 &5900 & 4.75799& $-$\\
0.004 & 0.746 & 5.4 & 2238.72 &5900 & 5.2892 & $-$\\
0.004 & 0.746 & 5.4 & 2238.72 &5800 & 5.62694 & 4.04213\\
0.008 & 0.742 & 5.6 & 1995.26 &5500 & 6.12775 & 6.1435\\
0.008 & 0.742 & 5.8 & 2238.72 &5500 & 6.60143 & 6.61745\\
0.004 & 0.746 & 6.2 & 3548.13 &5900 & 7.14526 & 5.09709\\
0.008 & 0.742 & 6.2 & 2818.38 &5400 & 8.22863 & 8.26111\\

\hline
\end{tabular}
}
\begin{tablenotes}
	\small
	\item $^{*}$Period is obtained from Set D.	        
\end{tablenotes}

\label{tab:Models}
\end{table*}

\begin{figure*}
\centering
\includegraphics[scale = 0.85]{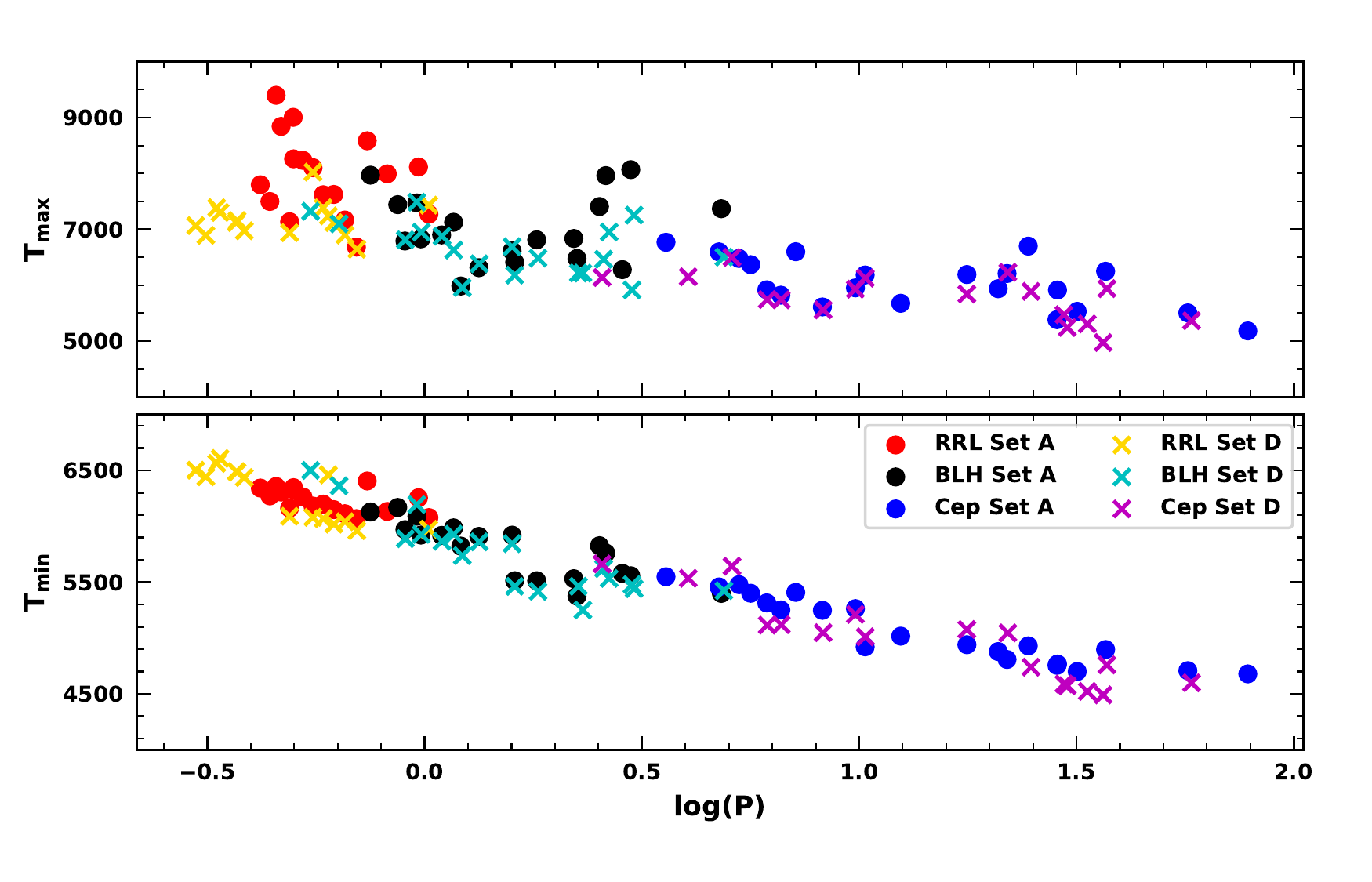}
\caption{The variation of temperature at maximum and minimum light with period for the computed models using convection sets A and D.}
\label{fig:period_temp} 
\end{figure*}

Several aspects of our theoretical approach have been outlined in the introduction and earlier sections and in earlier papers \citep{simon1993, kanbur1995, kanbur1996, kanbur2004b, kanbur2006, kanbur2007}. In summary, the properties of mean light PC relations are determined by the properties of PC relations at different phases during a pulsation cycle and at maximum and minimum light, these are determined by the interaction of the stellar photosphere and the HIF.

In an effort to understand the interaction of the HIF and the stellar photosphere during the pulsation cycle, we theoretically study the HIF-stellar photosphere distance \citep[see][]{kanbur2004b} at minimum and maximum light using the radial stellar pulsation code in MESA. We use MESA~r-$11701$ \citep{paxton2019} to compute RR~Lyrae, BL~Her and classical Cepheid models, the parameters of which are listed in Table~\ref{tab:Models}. We chose this pulsation code because it is a state-of-the-art pulsation code that is open source. Further it allows some flexibility in treatments of turbulent convection and opens up the possibility of using future detailed studies to constrain theories of turbulent convection in classical Cepheids and RR Lyraes. Here we show that in terms of light curves, it matches observations well. In common with many other pulsation codes \citep{marconi2015} it uses the diffusion approximation to model radiative transfer, though future plans include more sophisticated
treatments of radiative transfer.

The input parameters for these models ($Z$, $X$, $M/M_{\odot}$, $L/L_{\odot}$ and ${T_\text{eff}}$) have been taken from literature (see e.g. \citealt{marconi2015} for RR Lyraes, \citealt{marconi2007} for BL~Her models and Table~1 of \citealt{kanbur2004b} for classical Cepheids). We may choose to use any of the four sets of convection parameters as outlined in Table~4 of \citet{paxton2019}. Set A corresponds to the simplest convection model, set B adds radiative cooling, set C adds turbulent pressure and turbulent flux, and set D includes these effects simultaneously. In the present work, we choose to use the simplest (Set A) and the most complex (Set D) convection models for the same input parameters. A study of the effect of different sets of convection parameters on the light curve structures of the models is planned for the future. For our analysis, we proceed with only those models that have full-amplitude stable pulsations in the fundamental mode. Fourier analysis of these models shows considerably good match between the models and observations in the Galactic bulge, LMC and SMC for OGLE-IV RR~Lyraes, BL~Her stars and classical Cepheids with respect to their Fourier parameters and thus, their light curve structures.

\begin{table*}
\caption{The slopes and intercepts for period-temperature relations of the mathematical form $\log(T)=a\log(P)+b$ for RR~Lyrae, BL~Her and Cepheid models at minimum and maximum light.}
\centering
\begin{tabular}{c c c c c c c}
\hline\hline
Model & \multicolumn{3}{c}{Minimum light} & \multicolumn{3}{c}{Maximum light}\\
\hline \hline
\multicolumn{7}{c}{Convection set A}\\
\hline
& $a$ & $b$ & $\sigma$ & $a$ & $b$ & $\sigma$ \\
\hline
RRL & -0.032$\pm$0.014 & 3.787$\pm$0.004 & 0.006 & -0.1$\pm$0.082 & 3.877$\pm$0.021 & 0.037\\
BLH & -0.07$\pm$0.012 & 3.776$\pm$0.003 & 0.011 & 0.012$\pm$0.041 & 3.841$\pm$0.012 & 0.036\\
Cep & -0.062$\pm$0.005 & 3.775$\pm$0.006 & 0.008 & -0.052$\pm$0.016 & 3.839$\pm$0.019 & 0.026\\
\hline
\multicolumn{7}{c}{Convection set D}\\
\hline
& $a$ & $b$ & $\sigma$ & $a$ & $b$ & $\sigma$ \\
\hline
RRL & -0.092$\pm$0.016 & 3.768$\pm$0.006 & 0.009 & 0.013$\pm$0.034 & 3.859$\pm$0.012 & 0.018\\
BLH & -0.089$\pm$0.012 & 3.776$\pm$0.004 & 0.012 & -0.054$\pm$0.026 & 3.832$\pm$0.008 & 0.026\\
Cep & -0.076$\pm$0.008 & 3.784$\pm$0.01 & 0.013 & -0.048$\pm$0.016 & 3.816$\pm$0.019 & 0.023\\
\hline
\end{tabular}
\label{tab:Models_PT}
\end{table*}

\subsection{Theoretical period-temperature and period-colour relations at maximum and minimum light}
The variation of temperature with period for the computed models at minimum and maximum light is displayed in Fig.~\ref{fig:period_temp} and tabulated in Table~\ref{tab:Models_PT}. We find a relatively well-defined period-temperature relation at minimum light across different types of pulsating variable stars considered. However, there is a higher dispersion in the period-temperature relation for the same models at maximum light. Considering RR~Lyrae models, while the temperature range between the reddest and the bluest models at minimum light is $\sim$500K, it is as high as $\sim$2000K at maximum light. Also note that if we were to expand the period scale, we would find the period-temperature relation for RRab stars at minimum light to be relatively shallow, consistent with observations. From Table~\ref{tab:Models_PT} we find that a flat (or flatter) period-temperature relation at maximum light within 3$\sigma$ uncertainties exists for RR~Lyraes, BL~Her and Cepheid models using both sets of convection parameters while RRL models also have a flat period-temperature relation at minimum light using the convection set~A.

\begin{figure*}
\centering
\includegraphics[scale = 0.85]{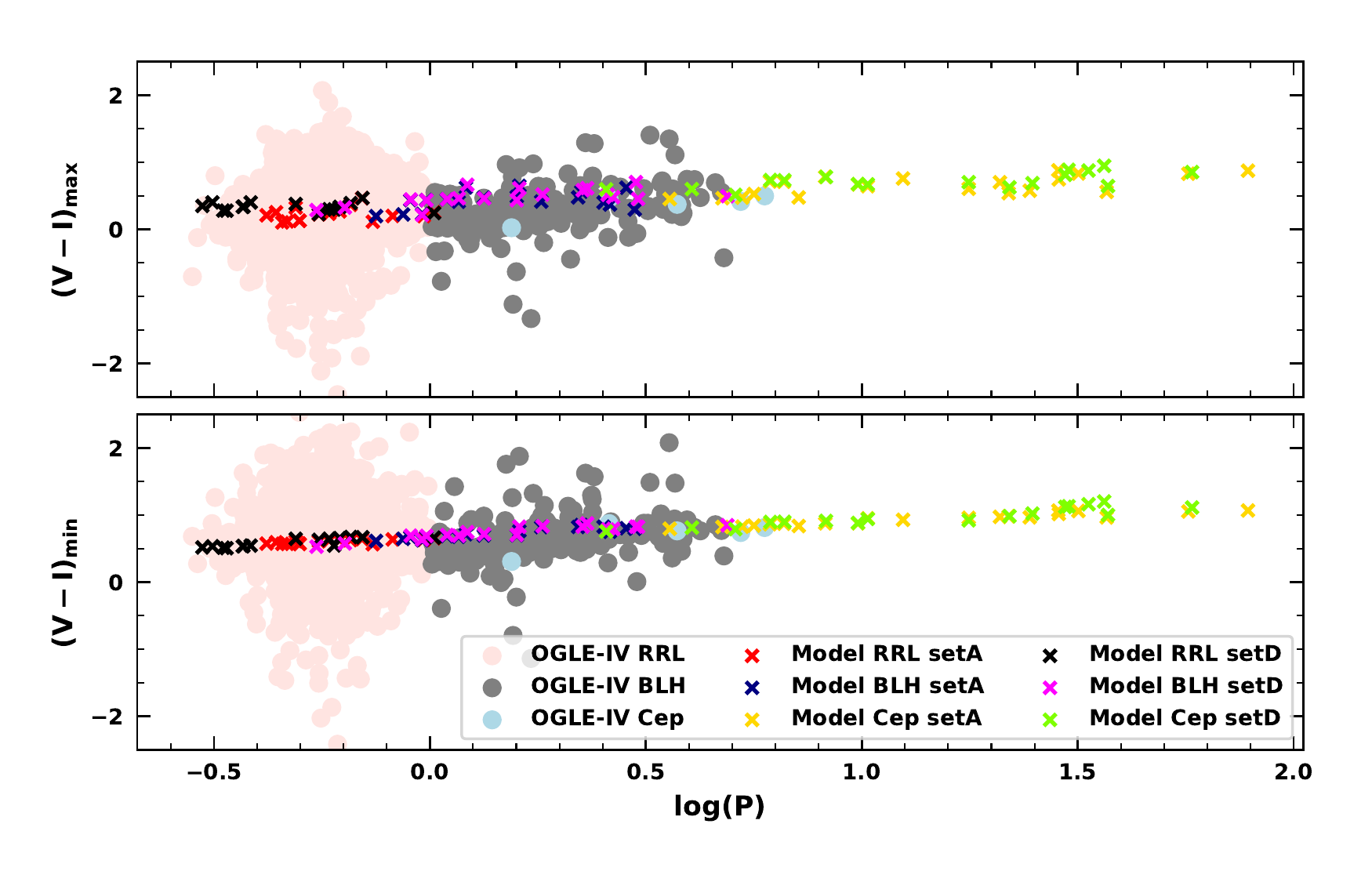}
\caption{A comparison of the theoretical period-colour relations for the computed models with observations from OGLE-IV RR~Lyraes, BL~Her stars and classical Cepheids in the Galactic bulge.}
\label{fig:PC_Bulge} 
\end{figure*}

\begin{figure*}
\centering
\includegraphics[scale = 0.85]{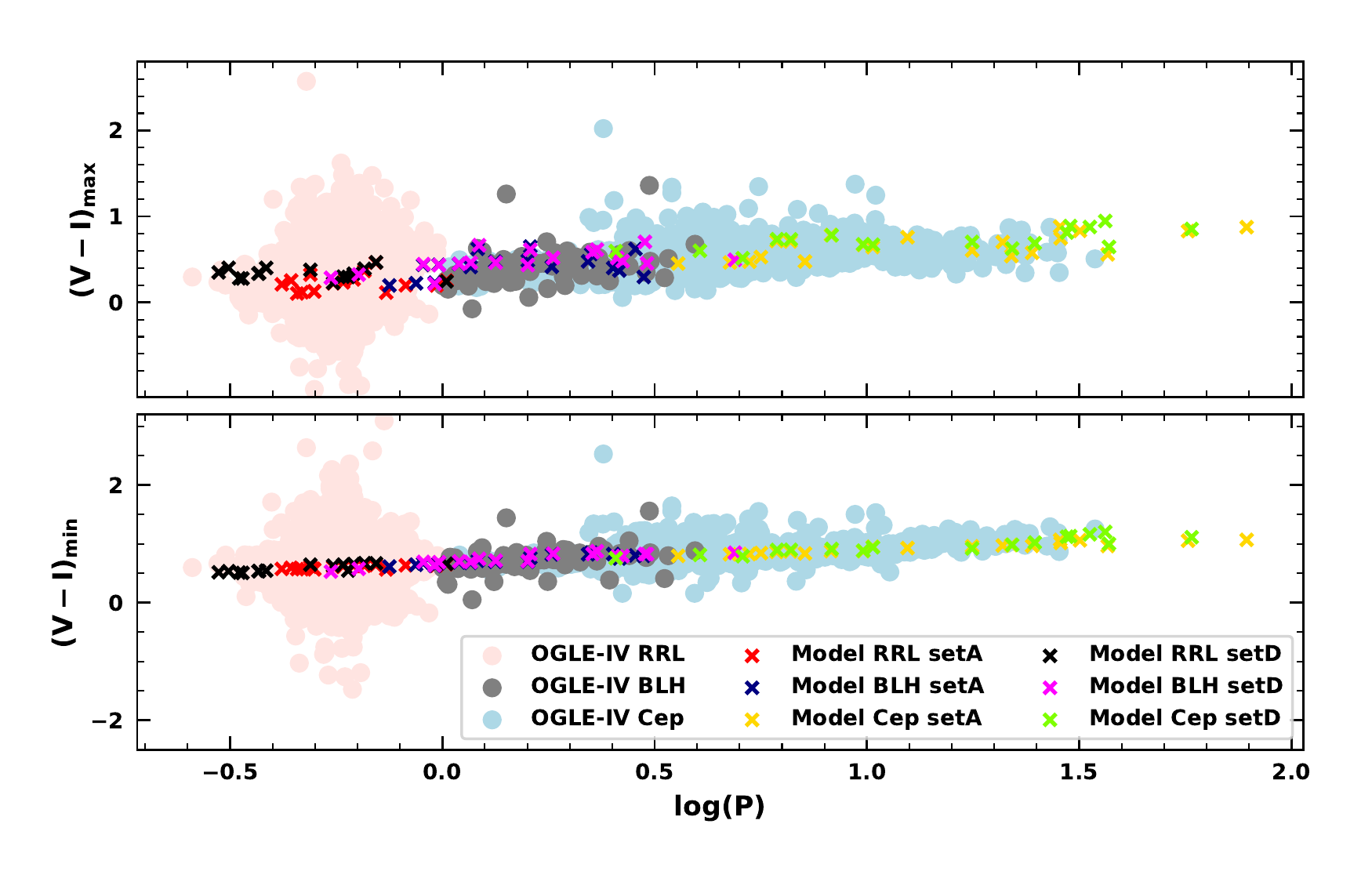}
\caption{Same as Fig.~\ref{fig:PC_Bulge} but for observations from LMC.}
\label{fig:PC_LMC} 
\end{figure*}

\begin{figure*}
\centering
\includegraphics[scale = 0.85]{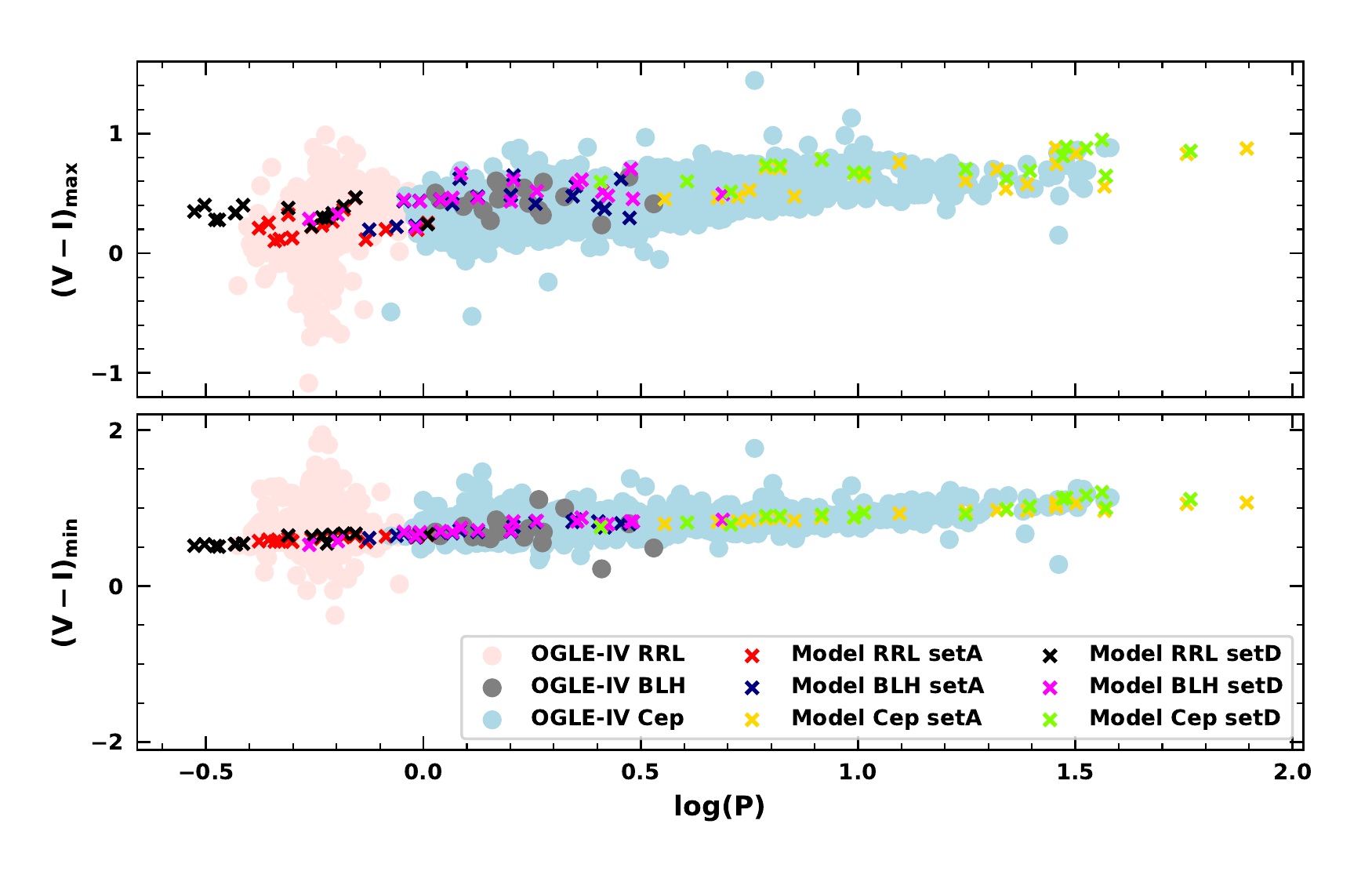}
\caption{Same as Fig.~\ref{fig:PC_Bulge} but for observations from SMC.}
\label{fig:PC_SMC} 
\end{figure*}

The PC and AC relations from the computed models are summarised in Table~\ref{tab:PCAC_models}. The PC slopes from RR~Lyrae and Cepheid models do not compare well with observations which show flat PC$_{\mathrm{min}}$ for RRab stars and flat PC$_{\mathrm{max}}$ for long-period classical Cepheids. However, we point here that the reason may be due to the fewer number of models used. A future study would involve computing a fine grid of models to rigorously constrain models that compare well with observations with respect to their PC slopes. For the purpose of this paper, we show the comparison of the theoretical PC relations for the computed models with observations from OGLE-IV RR~Lyraes, BL~Her stars and classical Cepheids in the Galactic bulge, LMC and SMC in Figs.~\ref{fig:PC_Bulge}, \ref{fig:PC_LMC} and \ref{fig:PC_SMC}, respectively. There is a good overlap in the PC relations between models and observations from the Galactic bulge, LMC and SMC at both minimum and maximum light. The models have consistent results with observations with respect to both Fourier parameters as well as colour. Subsequently, we use these models to probe the temperature and opacity profiles and to study the distance between the HIF and the stellar photosphere in an attempt to understand the microphysics of the outer envelopes of these pulsating variable stars.

\subsection{Temperature and opacity profiles}
\begin{figure*}
\centering
\includegraphics[scale = 0.85]{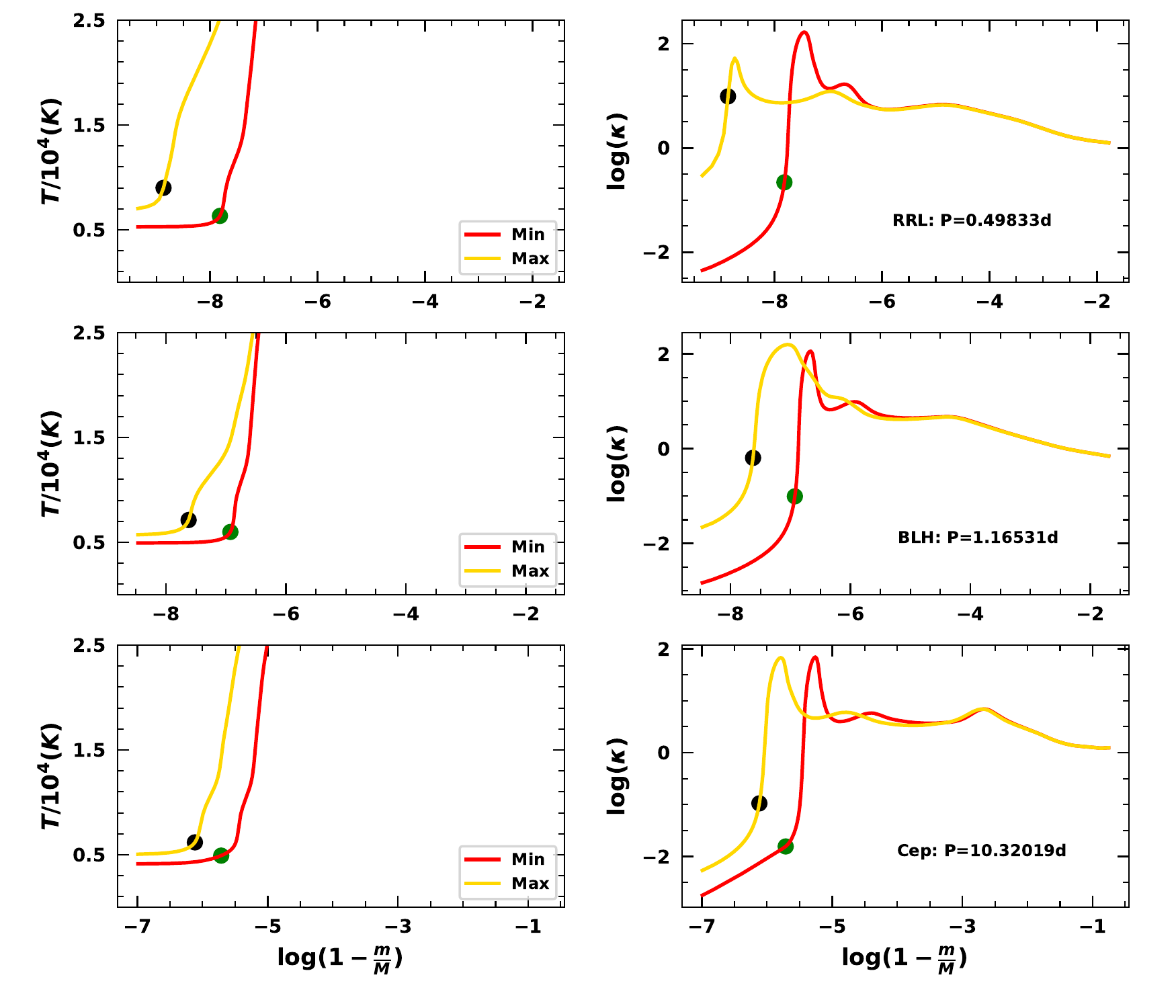}
\caption{The temperature and opacity profiles for RR~Lyrae, BL~Her and classical Cepheid models. The filled circles are the location of the photosphere at $\tau=2/3$ for each phase. Left panel: temperature profile. Right panel: opacity profile.}
\label{fig:profiles} 
\end{figure*}

Fig.~\ref{fig:profiles} presents the plots of the temperature and opacity profiles as a function of mass distribution for each of RR~Lyrae, BL~Her and classical Cepheid models. The mass distribution is measured by the quantity $Q=\log(1-M_r/M)$, where $M_r$ is the mass within radius, $r$ and $M$ is the total mass. The photosphere is defined as the zone with optical depth, $\tau=2/3$. From Fig.~\ref{fig:profiles}, we find that the photosphere (shown in filled circles) at maximum light lies further out in the mass distribution than at minimum light; this is expected as the star expands and contracts during its pulsation cycle. However, we observe that the photosphere at maximum light lies much further out in the mass distribution in the case of RR~Lyrae ($Q\sim-9$) and BL~Her ($Q\sim-7.5$) as compared to Cepheids ($Q\sim-6$). Both the temperature and opacity values increase as we move from Cepheids to BL~Her to RR~Lyraes at maximum and minimum light. The HIF and stellar photosphere can either be engaged or disengaged. The top two left panels of Fig.~\ref{fig:profiles} show an engaged HIF and photosphere while the bottom left panel depicts an example of where they are not engaged. In fact this panel shows an example of where the HIF and photosphere are engaged at one phase (maximum light) and disengaged at another phase (minimum light). When the HIF and photosphere are disengaged, the temperature of the photosphere and hence the colour of the star is more dependent on global stellar parameters and hence exhibits a relationship with period. When the HIF and photosphere are engaged at low temperature ($T<6300 {\textrm K}$), the temperature of the photosphere and hence the temperature of the photosphere is somewhat independent of density and hence the relation with period is reduced (e.g. RRab's at minimum light, Cepheids at maximum light). When the HIF and photosphere are engaged at higher temperature (RR Lyraes at maximum light, overtones), ionization equilibrium is more sensitive to temperature. In this situation, the temperature of the photosphere is the temperature at which hydrogen ionizes and this has a stronger dependency on period.

\subsection{HIF-Stellar Photosphere Distance}

\begin{figure*}
\centering
\includegraphics[scale = 0.85]{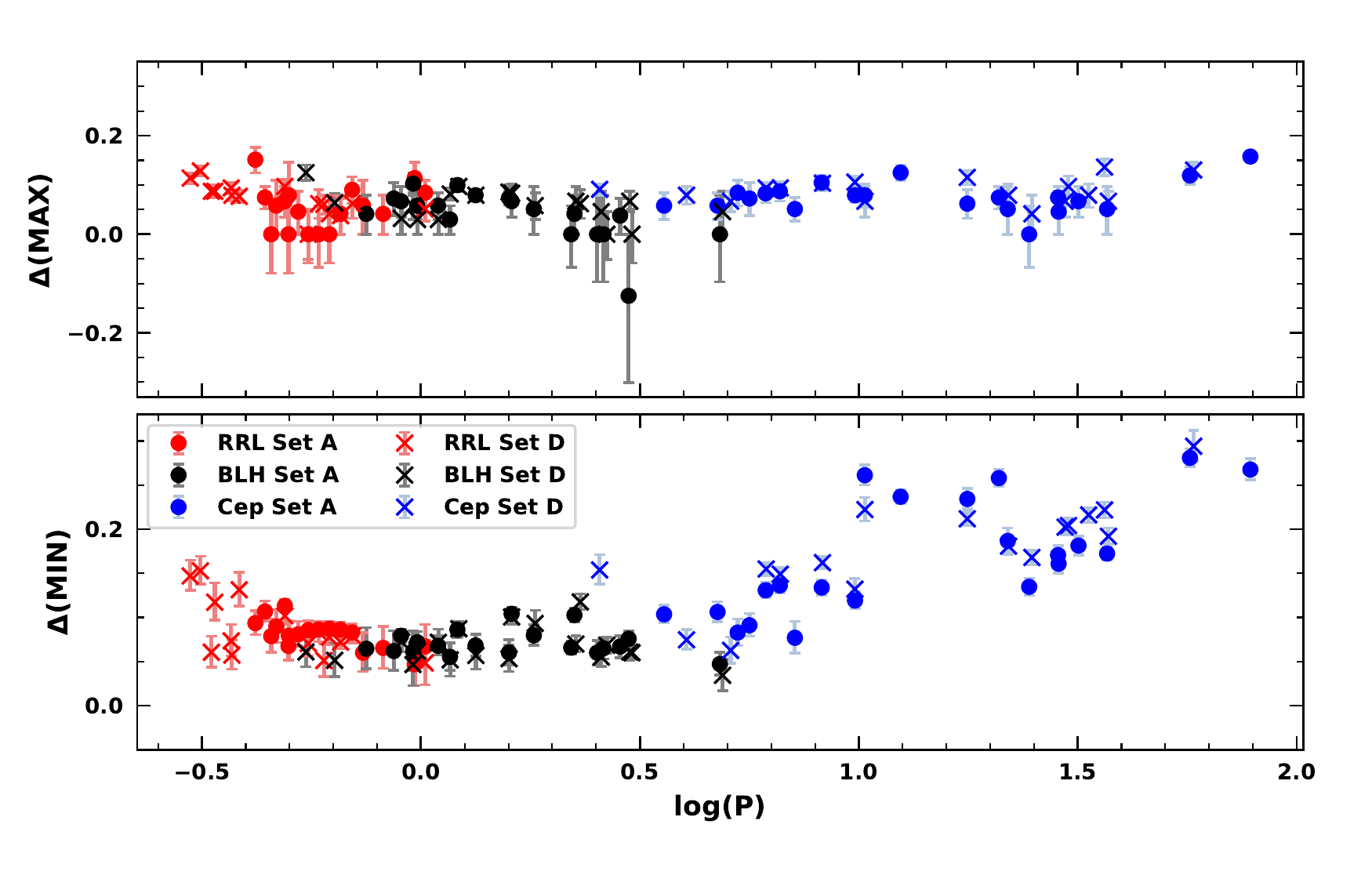}
\caption{The plots of distance ($\Delta$) between HIF and stellar photosphere as function of $\log(P)$ at maximum and minimum light. The error bars are estimated from the coarseness of the grid points around the location of HIF.}
\label{fig:HIF_Ph_distance} 
\end{figure*}

The HIF is defined as the zone with the steepest gradient in temperature. The distance between the stellar photosphere and HIF, $\Delta$ may then be defined in terms of $Q=\log(1-M_r/M)$. The interested reader is referred to Fig.~9 of \citet{kanbur2004b} for a clear, pictorial representation. Fig.~\ref{fig:HIF_Ph_distance} displays the distance between the HIF and stellar photosphere as a function of $\log(P)$ for a range of theoretical models from RR Lyraes to BL Hers to classical Cepheids. The error bars are estimated from the coarseness of the grid points around the location of HIF. The top and bottom panels represent the distance between the HIF and stellar photosphere at maximum/minimum light respectively. Firstly we note that at maximum light, the HIF and photosphere are always closer together: the distance between HIF and photosphere is always smaller for a given model at maximum light. For RR Lyraes, the HIF and photosphere are always engaged, but at minimum light, this occurs in a temperature regime for which the ionization of hydrogen is somewhat independent of temperature. Because the HIF and photosphere are engaged, temperature of the photosphere is just the same as the temperature at which hydrogen is ionized and this occurs at a temperature that is somewhat independent of period - hence we have flat/shallow PC relations at minimum light for RRab stars. At maximum light, the star is hotter, the HIF and photosphere are still engaged but in a regime that is more sensitive to temperature changes and hence the temperature of the photosphere, which is still the temperature at which hydrogen ionizes, is more sensitive to global stellar parameters and hence we have PC relation at maximum light for RRab stars. Because the RRc stars are hotter, they exhibit non-flat PC relations at maximum/minimum light.

Cepheids have higher $L/M$ ratios and are cooler than RR Lyraes and hence from the work of \citet{kanbur1995} and \citet{kanbur1996} the HIF lies further inside the mass distribution than in the case for RR Lyraes. The HIF and photosphere are disengaged at minimum light and hence the temperature of the photosphere and hence the colour at minimum light are dependent on stellar parameters and we have PC relation with a significant slope. The HIF and photosphere are engaged at maximum light at low temperature and hence we have flattish PC relation at maximum phase. We may regard BL Her stars as intermediate between Cepheids and RR Lyraes. At minimum light there is sufficient disengagement so that the temperature of the photosphere is somewhat dependent on global stellar parameters and hence period. At maximum light, the situation is similar to maximum light for RRab stars. BL Her stars are cooler and brighter than RR Lyraes. Our analysis from Section~\ref{sec:theory} and the work of \citet{kanbur1995} and \citet{kanbur1996} suggest that the HIF is further inside the mass distribution, thus increasing the distance between the HIF and stellar photosphere.

Fig.~\ref{fig:HIF_Ph_distance} is broadly consistent with this picture. The top panel shows a fairly constant distance between stellar photosphere and HIF at maximum light across variable star type. An interesting result to note from the top panel is the presence of models with zero or negative distance between the HIF and the stellar photosphere at maximum light - a situation where the photosphere ``climbs'' up the HIF. A detailed study of the $D$ to $R$ ionization front transition \citep[see][]{adams1979} in Cepheids and RR Lyraes may be useful to analyse this situation. The bottom panel (distance between stellar photosphere and HIF at minimum light) displays a sharp jump in the distance at a period of about $\log(P) \approx 0.5$ followed by an approximately linear increase with period. 

WVir stars are highly adiabatic and this poses some problems for existing stellar pulsation codes - hence we have not used MESA-RSP to model WVir stars. However, since they occupy a similar region of the HR Diagram as classical Cepheids, our contention is that the HIF-stellar photosphere interaction will be similar to that in Cepheids. Figs.~\ref{fig:hif-rrl} and \ref{fig:hif-agb} provide strong evidence that as stars move off the HB and upward along the AGB, the relative positions of the HIF and stellar photosphere change in such a way as to make this plausible.

\begin{table*}
\caption{The slopes and intercepts for PC and AC relations for the RRab, BL~Her and classical Cepheid models.}
\centering
\begin{tabular}{c c c c c c c c}
\hline\hline
Model & Phase & \multicolumn{3}{c}{PC} & \multicolumn{3}{c}{AC}\\
\hline \hline
& & Slope & Intercept & $\sigma$ & Slope & Intercept & $\sigma$\\
\hline
\multicolumn{8}{c}{Convection set A}\\ 
\hline
RRab & max & 0.184$\pm$0.206 & 0.265$\pm$0.052 & 0.091 & -0.428$\pm$0.032 & 0.635$\pm$0.031 & 0.026\\
 & min & 0.181$\pm$0.045 & 0.646$\pm$0.011 & 0.02 & -0.072$\pm$0.03 & 0.673$\pm$0.029 & 0.024\\
BLH & max & 0.243$\pm$0.15 & 0.376$\pm$0.039 & 0.113 & -0.426$\pm$0.058 & 0.733$\pm$0.046 & 0.058\\
 & min & 0.346$\pm$0.044 & 0.668$\pm$0.012 & 0.034 & -0.016$\pm$0.076 & 0.74$\pm$0.06 & 0.075\\
Cep & max & 0.252$\pm$0.089 & 0.356$\pm$0.122 & 0.101 & -0.497$\pm$0.116 & 1.1$\pm$0.098 & 0.082\\
 & min & 0.205$\pm$0.029 & 0.704$\pm$0.039 & 0.032 & -0.094$\pm$0.102 & 1.054$\pm$0.086 & 0.072\\

\hline
\multicolumn{8}{c}{Convection set D}\\ 
\hline
RRab & max & -0.036$\pm$0.117 & 0.318$\pm$0.041 & 0.063 & -0.14$\pm$0.067 & 0.421$\pm$0.046 & 0.054\\
 & min & 0.36$\pm$0.064 & 0.704$\pm$0.022 & 0.034 & 0.203$\pm$0.055 & 0.458$\pm$0.038 & 0.045\\
BLH & max & 0.289$\pm$0.097 & 0.426$\pm$0.03 & 0.097 & -0.35$\pm$0.122 & 0.71$\pm$0.083 & 0.098\\
 & min & 0.343$\pm$0.042 & 0.678$\pm$0.013 & 0.042 & 0.001$\pm$0.116 & 0.743$\pm$0.079 & 0.094\\
Cep & max & 0.213$\pm$0.079 & 0.468$\pm$0.107 & 0.087 & -0.092$\pm$0.17 & 0.826$\pm$0.147 & 0.113\\
 & min & 0.316$\pm$0.056 & 0.605$\pm$0.076 & 0.062 & 0.167$\pm$0.183 & 0.881$\pm$0.158 & 0.121\\
\hline
\end{tabular}
\label{tab:PCAC_models}
\end{table*}

\section{Discussion and Conclusions}
\label{sec:results}

We have analysed the largest available dataset of T2Cs in the Bulge, LMC and SMC from the OGLE-IV survey for the first time. We found that there exists a flat PC slope at maximum light for W~Vir stars in Bulge, LMC and SMC within 3$\sigma$ uncertainties. This is similar to the flat PC$_\textrm{max}$ slope observed in long period classical Cepheids from present work and \citet{bhardwaj2014}. Also, the PC$_\textrm{min}$ slope of W~Vir stars is much greater than the slope at maximum light. We observed a flat but slightly negative PC$_\textrm{min}$ slope in BL~Her stars from SMC ($-0.251 \pm 0.339$) similar to that obtained for RR~Lyrae stars from SMC at minimum light \citep{das2018}. The slopes of the PC relations for RV~Tau stars in Bulge, LMC and SMC are flat within 3$\sigma$ uncertainties at minimum and maximum light. However, there is high dispersion in all the cases. Using the statistical $F$-test, we find a break in the PC relations at a period of 10 days for classical Cepheids in LMC and SMC from OGLE-IV at both minimum and maximum light, similar to earlier results from \citet{kanbur2004a} and \citet{bhardwaj2014}. There exists a flat PC$_\textrm{max}$ within 3$\sigma$ uncertainties and a significantly sloped PC$_\textrm{min}$ for the longer period classical Cepheids in both LMC and SMC. We have also carried out $t$-test to study the equivalence of PC slopes in LMC and SMC across a broad spectrum of variable star types and found the results to statistically support the above mentioned results.

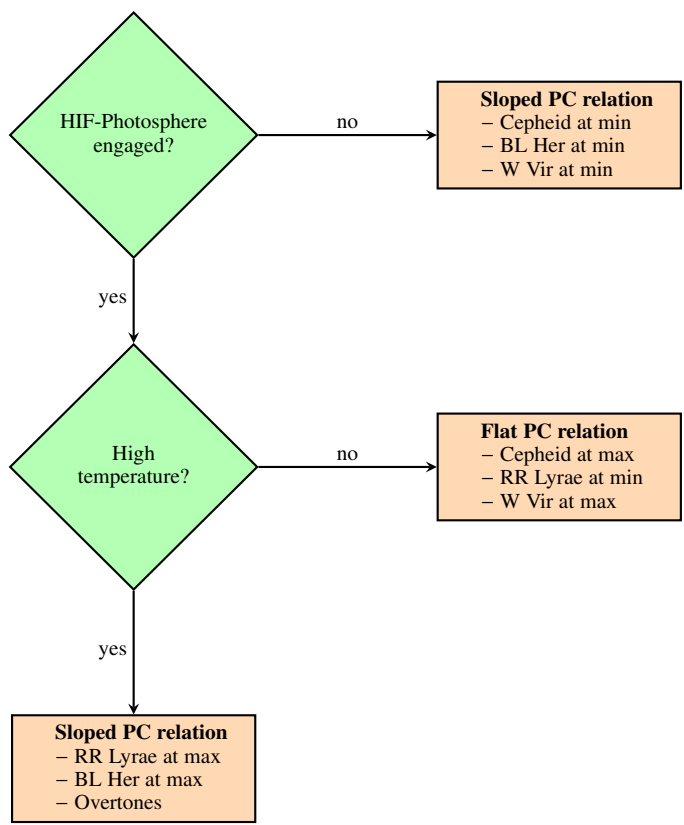
\begin{figure}
\centering
\begin{tikzpicture}[node distance=5cm, thick,scale=0.8, every node/.style={scale=0.8}]
\node (dec1) [decision, yshift=-0.5cm] {\parbox{3cm}{\centering \large HIF-Photosphere \\ engaged?}};
\node (pro1) [process, right of=dec1, xshift=2cm] {\large \begin{tabular}{l} {\bf Sloped PC relation} \\ $-$ Cepheid at min \\ $-$ BL~Her at min \\ $-$ W~Vir at min\end{tabular}};
\node (dec2) [decision, below of=dec1, yshift=-0.5cm] {\parbox{3cm}{\centering \large High \\ temperature?}};
\node (pro2) [process, right of=dec2, xshift=2cm] {\large \begin{tabular}{l} {\bf Flat PC relation} \\ $-$ Cepheid at max \\ $-$ RR~Lyrae at min \\  $-$ W~Vir at max\end{tabular}};
\node (pro3) [process, below of=dec2] {\large \begin{tabular}{l} {\bf Sloped PC relation} \\ $-$ RR~Lyrae at max \\ $-$ BL~Her at max \\  $-$ Overtones\end{tabular}};
\draw [arrow] (dec1) -- node[anchor=east] {\large yes} (dec2);
\draw [arrow] (dec1) -- node[anchor=south] {\large no} (pro1);
\draw [arrow] (dec2) -- node[anchor=east] {\large yes} (pro3);
\draw [arrow] (dec2) -- node[anchor=south] {\large no} (pro2);
\end{tikzpicture}
\caption{A summary of the HIF-stellar photosphere interaction theory explaining the different behaviour of PC relations across a broad spectrum of variable star types at minimum and maximum light.}
\label{fig:flowchart} 
\end{figure}

We also computed a few RR~Lyrae, BL~Her and classical Cepheid models using the radial stellar pulsation code in MESA to theoretically study the distance between HIF and stellar photosphere at minimum and maximum light. We found that RRL models have a flat period-temperature relation at minimum and maximum light within 3$\sigma$ uncertainties using both sets of convection parameters A and D and a flat period-temperature relation at maximum light within 3$\sigma$ uncertainties exists for BL~Her and Cepheid models using both sets of convection parameters. The period-temperature relation for the models has a higher dispersion at maximum light than at minimum light. We also found the models to have consistent results with observations with respect to both Fourier parameters as well as colour. The HIF and stellar photosphere are engaged and co-moving at maximum light for classical Cepheids, which results in a flat PC slope at maximum light. However, they are disengaged at minimum light and we observe a significant slope in PC$_\textrm{min}$. For RR~Lyraes, the HIF and stellar photosphere are engaged during both minimum and maximum light. However, at maximum light, the temperature at which hydrogen starts to ionize appreciably is in a range where Saha ionization equilibrium is much more sensitive to temperature than at the temperatures associated with minimum light, even though the range of densities are similar. This results in a flat PC$_\textrm{min}$ slope but a significant slope in PC$_\textrm{max}$. BL Her stars are cooler and brighter than RR Lyraes; thus we suggest that the HIF is further inside the mass distribution, thereby increasing the distance between the HIF and stellar photosphere. The models computed using MESA are broadly consistent with this picture. We have not computed W~Vir stars using MESA; however, similar locations of W~Vir stars and classical Cepheids on the HR diagram and similar behaviour of PC relations from observations for both suggest similar HIF and stellar photosphere interaction. Thus, in this work, we have incorporated the PC and AC relations for RR~Lyraes, BL~Her, W~Vir and classical~Cepheid stars in a single unifying theory that involves the interaction of the hydrogen ionization front and stellar photosphere and the theory of stellar evolution. Fig.~\ref{fig:flowchart} summarizes the main results in the form of a flowchart. An exception to these results are the flat PC relations for BL~Her stars at both minimum and maximum light and for W~Vir stars at minimum light in the SMC. However, we note here that the statistical sample of stars in the SMC is small ($\sim 20$ BL~Her and $\sim 15$ W~Vir stars).

\citet{paxton2019} have outlined the physical and numerical assumptions incorporated in the MESA-RSP code and demonstrated that it produces stable, multi-wavelength light curve models for a broad spectrum of variable stars. To evaluate possible systematics in MESA-RSP models due to the input physics such as the equation of state, opacities or color-temperature transformations, it is important to compare these with different pulsation models in the literature.
Since MESA-RSP has just been published \citep{paxton2019}, detailed comparisons between pulsation results with this code and others in the literature have not yet been 
carried out. We note that comparisons between observations and data for a wide class of radially pulsating stars were reasonable \citep{paxton2019}.
MESA-RSP follows the work of \cite{smolec2008} in its treatment of stellar pulsation and uses a particular theory of
turbulent convection \citep{kuhfuss1986}. Other similar codes \citep{marconi2015} use the convection formulation outlined in \citet{stellingwerf1982a, stellingwerf1982b}. MESA-RSP affords the
possibility of varying the convection parameters used. Using turbulent convection parameter sets A and D \citep{paxton2019} suggest that our results are robust to this. In common
with other pulsation codes in the literature, MESA-RSP uses the diffusion approximation to account for radiative transfer. Relaxing this assumption is certainly a project for the future but we do not anticipate that a more detailed radiative transfer will affect the general character of the results presented here.
Previously, \citet{simon1993} and \citet{kanbur1996} used purely radiative models, whilst \citet{kanbur2007} and \citet{das2018} used convection theories from \citet{kuhfuss1986} and \citet{stellingwerf1982a}, respectively and found similar qualitative results regarding the behaviour of PC relations of classical Cepheids and RR~Lyraes. While a detailed comparison of input physics in different pulsation models is beyond the scope of this paper, we are computing a fine grid of Cepheid and RR~Lyrae models covering the entire parameter space using MESA-RSP, which will be used to explore possible systematics and constrain pulsation models in the future.

\section*{Acknowledgements}

The authors thank the referee for useful comments and suggestions that improved the quality of the manuscript. SD acknowledges the INSPIRE Senior Research Fellowship vide Sanction Order No. DST/INSPIRE Fellowship/2016/IF160068 under the INSPIRE Program from the Department of Science \& Technology, Government of India. HPS and SMK thank the Indo-US Science and Technology Forum for funding the Indo-US virtual joint networked centre on ``Theoretical analyses of variable star light curves in the era of large surveys''. SD acknowledges the travel support provided by SERB, Government of India vide file number ITS/2019/004781 to attend the RR Lyrae/Cepheid 2019 Conference ``Frontiers of Classical Pulsators: Theory and Observations'', USA where the authors had useful discussions. Funding for the Stellar Astrophysics Centre is provided by The Danish National Research Foundation (Grant agreement no.: DNRF106). AB acknowledges research grant \#11850410434 awarded by the National Natural Science Foundation of China through the Research Fund for International Young Scientists, China Postdoctoral General Grant (2018M640018), and Peking University Strategic Partnership Fund awarded to Peking-Tokyo Joint-collaboration on Research in Astronomy and Astrophysics. BM and NP acknowledge IUSSTF for supporting travel to University of Delhi and AC and RJ thank SUNY Oswego and the NSF STEP grant. The authors acknowledge the use of High Performance Computing facility Pegasus at IUCAA, Pune and the following software used in this project: MESA~r$10108$ and MESA~r$11701$ \citep{paxton2011,paxton2013,paxton2015,paxton2018,paxton2019}. 

\bibliographystyle{mnras}

\appendix
\section{Statistical $t$-test tables}
\label{sec:appendix}
The results from the statistical $t$-test to compare PC slopes in the LMC and SMC across a broad spectrum of variable star types are summarised in Tables~\ref{tab:ttest_LMC} and \ref{tab:ttest_SMC}, respectively and discussed in Sec.~\ref{sec:t-test}.

\begin{table*}
\caption{Results of $t$-test on PC slopes in the LMC$^{*}$.}
\centering
\resizebox{\textwidth}{!}{
\begin{tabular}{c c c c c c c c c c}
\hline\hline
& BL~Her & W~Vir & pW~Vir & RV~Tau & T2C & RRab & Cep${_S}$ & Cep${_L}$ & Cep${_{all}}$\\
\hline
\multicolumn{10}{c}{Minimum light}\\
\hline
BL~Her & ... & ... & ... & ... & ...  & ...  & ...  & ...  & ... \\
W~Vir &(0.055,1.975,0.478) & ...  & ...  & ...  & ...  & ...  & ...  & ...  & ... \\
pW~Vir &(0.508,1.987,0.306) &(0.500,1.982,0.309) & ...  & ...  & ...  & ...  & ...  & ...  & ... \\
RV Tau &(1.485,1.981,0.070) &(1.487,1.977,0.070) &(1.068,1.994,0.145) & ...  & ...  & ...  & ...  & ...  & ... \\
T2C &{\bf(3.787,1.968,0.000)} &{\bf(5.045,1.967,0.000)} &(1.015,1.969,0.156) &(0.657,1.968,0.256) & ...  & ...  & ...  & ...  & ... \\
RRab &{\bf(4.321,1.960,0.000)} &{\bf(5.877,1.960,0.000)} &(1.200,1.960,0.115) &(0.558,1.960,0.288) &(1.546,1.960,0.061) & ...  & ...  & ...  & ... \\
Cep${_S}$ &{\bf(3.575,1.961,0.000)} &{\bf(4.878,1.961,0.000)} &(0.889,1.961,0.187) &(0.730,1.961,0.233) &(1.216,1.961,0.112) &{\bf(4.825,1.960,0.000)} & ...  & ...  & ... \\
Cep${_L}$ &(0.399,1.973,0.345) &(0.421,1.972,0.337) &(0.343,1.978,0.366) &(1.399,1.975,0.082) &{\bf(4.765,1.967,0.000)} &{\bf(5.650,1.960,0.000)} &{\bf(4.588,1.961,0.000)} & ...  & ... \\
Cep${_{all}}$ &{\bf(3.114,1.961,0.001)} &{\bf(4.242,1.961,0.000)} &(0.703,1.961,0.241) &(0.832,1.961,0.203) &{\bf(2.991,1.961,0.001)} &{\bf(7.991,1.960,0.000)} &{\bf(4.761,1.961,0.000)} &{\bf(3.909,1.961,0.000)} & ... \\
\hline
\multicolumn{10}{c}{Mean light}\\
\hline
BL~Her & ... & ... & ... & ... & ...  & ...  & ...  & ...  & ... \\
W~Vir &{\bf(4.523,1.976,0.000)} & ...  & ...  & ...  & ...  & ...  & ...  & ...  & ... \\
pW~Vir &(0.271,1.988,0.394) &(1.567,1.982,0.060) & ...  & ...  & ...  & ...  & ...  & ...  & ... \\
RV Tau &{\bf(2.309,1.981,0.011)} &(1.002,1.978,0.159) &{\bf(1.774,1.994,0.040)} & ...  & ...  & ...  & ...  & ...  & ... \\
T2C &{\bf(5.878,1.968,0.000)} &(0.087,1.967,0.465) &(1.598,1.969,0.056) &(1.039,1.968,0.150) & ...  & ...  & ...  & ...  & ... \\
RRab &(0.277,1.960,0.391) &{\bf(6.690,1.960,0.000)} &(0.369,1.960,0.356) &{\bf(2.420,1.960,0.008)} &{\bf(17.059,1.960,0.000)} & ...  & ...  & ...  & ... \\
Cep${_S}$ &{\bf(5.811,1.961,0.000)} &(0.412,1.961,0.340) &(1.510,1.961,0.066) &(1.108,1.961,0.134) &(0.887,1.961,0.188) &{\bf(25.542,1.960,0.000)} & ...  & ...  & ... \\
Cep${_L}$ &{\bf(1.905,1.973,0.029)} &{\bf(2.544,1.972,0.006)} &(0.522,1.977,0.301) &{\bf(1.741,1.974,0.042)} &{\bf(3.253,1.967,0.001)} &{\bf(2.899,1.960,0.002)} &{\bf(3.079,1.961,0.001)} & ...  & ... \\
Cep${_{all}}$ &{\bf(5.528,1.961,0.000)} &(0.733,1.961,0.232) &(1.416,1.961,0.078) &(1.174,1.961,0.120) &{\bf(1.823,1.961,0.034)} &{\bf(26.072,1.960,0.000)} &{\bf(1.782,1.961,0.037)} &{\bf(2.789,1.961,0.003)} & ... \\
\hline
\multicolumn{10}{c}{Maximum light}\\
\hline
BL~Her & ... & ... & ... & ... & ...  & ...  & ...  & ...  & ... \\
W~Vir &{\bf(3.538,1.975,0.000)} & ...  & ...  & ...  & ...  & ...  & ...  & ...  & ... \\
pW~Vir &(0.979,1.986,0.165) &{\bf(3.325,1.982,0.001)} & ...  & ...  & ...  & ...  & ...  & ...  & ... \\
RV Tau &{\bf(1.987,1.980,0.025)} &(0.170,1.978,0.433) &{\bf(2.397,1.995,0.010)} & ...  & ...  & ...  & ...  & ...  & ... \\
T2C &(1.619,1.968,0.053) &{\bf(3.409,1.967,0.000)} &{\bf(2.020,1.969,0.022)} &(1.359,1.968,0.088) & ...  & ...  & ...  & ...  & ... \\
RRab &{\bf(11.648,1.960,0.000)} &{\bf(19.370,1.960,0.000)} &{\bf(5.380,1.960,0.000)} &{\bf(8.460,1.960,0.000)} &{\bf(47.812,1.960,0.000)} & ...  & ...  & ...  & ... \\
Cep${_S}$ &(0.685,1.961,0.247) &{\bf(4.637,1.961,0.000)} &(1.496,1.961,0.067) &{\bf(1.877,1.961,0.030)} &{\bf(3.601,1.961,0.000)} &{\bf(57.780,1.960,0.000)} & ...  & ...  & ... \\
Cep${_L}$ &{\bf(3.922,1.973,0.000)} &(0.457,1.972,0.324) &{\bf(3.585,1.977,0.000)} &(0.431,1.974,0.334) &{\bf(3.963,1.967,0.000)} &{\bf(19.626,1.960,0.000)} &{\bf(5.178,1.961,0.000)} & ...  & ... \\
Cep${_{all}}$ &(1.410,1.961,0.079) &{\bf(3.798,1.961,0.000)} &{\bf(1.898,1.961,0.029)} &(1.495,1.961,0.068) &(0.962,1.961,0.168) &{\bf(67.469,1.960,0.000)} &{\bf(4.085,1.961,0.000)} &{\bf(4.358,1.961,0.000)} & ... \\
\hline
\end{tabular}
}
\begin{tablenotes}
	\small
	\item $^{*}$Values in parentheses are (T, $\mathrm{t_{\alpha/2,\nu}}$, p-value). The bold-faced entries indicate that the null hypothesis of the equivalent PC slopes can be rejected.	        
\end{tablenotes}
\label{tab:ttest_LMC}
\end{table*}

\begin{table*}
\caption{Results of $t$-test on PC slopes in the SMC$^{*}$.}
\centering
\resizebox{\textwidth}{!}{
\begin{tabular}{c c c c c c c c c c}
\hline\hline
& BL~Her & W~Vir & pW~Vir & RV~Tau & T2C & RRab & Cep${_S}$ & Cep${_L}$ & Cep${_{all}}$\\
\hline
\multicolumn{10}{c}{Minimum light}\\
\hline
BL~Her & ... & ... & ... & ... & ...  & ...  & ...  & ...  & ... \\
W~Vir &(1.682,2.040,0.051) & ...  & ...  & ...  & ...  & ...  & ...  & ...  & ... \\
pW~Vir &(1.404,2.069,0.087) &(0.449,2.101,0.329) & ...  & ...  & ...  & ...  & ...  & ...  & ... \\
RV Tau &(0.213,2.056,0.416) &(1.509,2.080,0.073) &(1.416,2.160,0.090) & ...  & ...  & ...  & ...  & ...  & ... \\
T2C &(1.288,1.995,0.101) &(1.278,1.998,0.103) &(0.787,2.004,0.217) &(1.204,2.002,0.117) & ...  & ...  & ...  & ...  & ... \\
RRab &(0.383,1.961,0.351) &{\bf(4.082,1.961,0.000)} &(1.447,1.961,0.074) &(0.541,1.961,0.294) &{\bf(4.758,1.961,0.000)} & ...  & ...  & ...  & ... \\
Cep${_S}$ &(1.315,1.961,0.094) &(1.401,1.961,0.081) &(0.785,1.961,0.216) &(1.220,1.961,0.111) &(0.060,1.961,0.476) &{\bf(8.958,1.960,0.000)} & ...  & ...  & ... \\
Cep${_L}$ &{\bf(2.247,1.981,0.013)} &(1.395,1.982,0.083) &(0.093,1.984,0.463) &{\bf(1.908,1.983,0.030)} &{\bf(4.061,1.976,0.000)} &{\bf(9.331,1.961,0.000)} &{\bf(5.474,1.961,0.000)} & ...  & ... \\
Cep${_{all}}$ &(1.399,1.961,0.081) &(1.140,1.961,0.127) &(0.725,1.961,0.234) &(1.282,1.961,0.100) &(0.574,1.961,0.283) &{\bf(9.813,1.960,0.000)} &{\bf(3.626,1.960,0.000)} &{\bf(5.002,1.961,0.000)} & ... \\
\hline
\multicolumn{10}{c}{Mean light}\\
\hline
BL~Her & ... & ... & ... & ... & ...  & ...  & ...  & ...  & ... \\
W~Vir &(0.828,2.040,0.207) & ...  & ...  & ...  & ...  & ...  & ...  & ...  & ... \\
pW~Vir &(1.408,2.069,0.086) &(1.075,2.101,0.148) & ...  & ...  & ...  & ...  & ...  & ...  & ... \\
RV Tau &(0.221,2.056,0.413) &(0.780,2.080,0.222) &(1.360,2.160,0.098) & ...  & ...  & ...  & ...  & ...  & ... \\
T2C &(1.301,1.995,0.099) &(0.890,1.998,0.188) &(0.865,2.004,0.195) &(1.056,2.002,0.148) & ...  & ...  & ...  & ...  & ... \\
RRab &{\bf(3.571,1.961,0.000)} &{\bf(6.028,1.961,0.000)} &(0.411,1.961,0.341) &{\bf(2.434,1.961,0.007)} &{\bf(10.733,1.961,0.000)} & ...  & ...  & ...  & ... \\
Cep${_S}$ &(1.461,1.961,0.072) &(1.284,1.961,0.100) &(0.791,1.961,0.215) &(1.146,1.961,0.126) &(0.733,1.961,0.232) &{\bf(21.038,1.960,0.000)} & ...  & ...  & ... \\
Cep${_L}$ &{\bf(2.549,1.981,0.006)} &{\bf(3.408,1.982,0.000)} &(0.125,1.984,0.450) &{\bf(1.837,1.983,0.035)} &{\bf(4.052,1.976,0.000)} &{\bf(3.367,1.961,0.000)} &{\bf(4.417,1.961,0.000)} & ...  & ... \\
Cep${_{all}}$ &(1.544,1.961,0.061) &(1.474,1.961,0.070) &(0.745,1.961,0.228) &(1.196,1.961,0.116) &(1.172,1.961,0.121) &{\bf(20.419,1.960,0.000)} &{\bf(2.526,1.960,0.006)} &{\bf(4.114,1.961,0.000)} & ... \\
\hline
\multicolumn{10}{c}{Maximum light}\\
\hline
BL~Her & ... & ... & ... & ... & ...  & ...  & ...  & ...  & ... \\
W~Vir &(0.574,2.040,0.285) & ...  & ...  & ...  & ...  & ...  & ...  & ...  & ... \\
pW~Vir &(1.346,2.069,0.096) &{\bf(1.810,2.101,0.044)} & ...  & ...  & ...  & ...  & ...  & ...  & ... \\
RV Tau &(0.473,2.056,0.320) &(0.823,2.080,0.210) &(0.551,2.160,0.295) & ...  & ...  & ...  & ...  & ...  & ... \\
T2C &(0.967,1.995,0.168) &{\bf(1.942,1.998,0.028)} &(0.963,2.004,0.170) &(0.051,2.002,0.480) & ...  & ...  & ...  & ...  & ... \\
RRab &{\bf(9.782,1.961,0.000)} &{\bf(11.910,1.961,0.000)} &{\bf(4.618,1.961,0.000)} &{\bf(4.252,1.961,0.000)} &{\bf(32.049,1.961,0.000)} & ...  & ...  & ...  & ... \\
Cep${_S}$ &(1.607,1.961,0.054) &{\bf(2.695,1.961,0.004)} &(0.580,1.961,0.281) &(0.249,1.961,0.402) &{\bf(3.011,1.961,0.001)} &{\bf(40.785,1.960,0.000)} & ...  & ...  & ... \\
Cep${_L}$ &(0.769,1.981,0.222) &{\bf(1.667,1.982,0.049)} &(1.052,1.983,0.148) &(0.131,1.983,0.448) &(0.421,1.976,0.337) &{\bf(22.623,1.961,0.000)} &{\bf(2.221,1.961,0.013)} & ...  & ... \\
Cep${_{all}}$ &(1.524,1.961,0.064) &{\bf(2.601,1.961,0.005)} &(0.632,1.961,0.264) &(0.209,1.961,0.417) &{\bf(2.642,1.961,0.004)} &{\bf(41.800,1.960,0.000)} &(1.170,1.960,0.121) &{\bf(1.995,1.961,0.023)} & ... \\
\hline
\end{tabular}
}
\begin{tablenotes}
	\small
	\item $^{*}$Values in parentheses are (T, $\mathrm{t_{\alpha/2,\nu}}$, p-value). The bold-faced entries indicate that the null hypothesis of the equivalent PC slopes can be rejected.	        
\end{tablenotes}
\label{tab:ttest_SMC}
\end{table*}

\label{lastpage}

\end{document}